\documentclass[12pt,a4paper]{article}
\usepackage{amsmath}
\usepackage{amssymb}
\usepackage[hidelinks]{hyperref}
\usepackage{graphicx,epsfig}
\usepackage[mathscr]{eucal}
\usepackage{subcaption}
\usepackage{color} 
\usepackage[numbers,sort&compress]{natbib}
\usepackage[english]{babel}

\usepackage{tabularx}

\usepackage{setspace}
\usepackage{xspace}

\usepackage{multirow}

\newcommand\as{\alpha_s}
 
\def\to{\rightarrow}

\def\msbar{{\overline {\rm MS}}}

\def\pT{p_T}

\def\citere#1{\mbox{Ref.\,\cite{#1}}}
\def\citeres#1{\mbox{Refs.\,\cite{#1}}}

\newcommand{\eqn}[1]{Eq.\,(\ref{#1})}

\newcommand{\fig}[1]{Figure\,\ref{#1}}
\newcommand{\figs}[1]{Figures\,\ref{#1}}
\newcommand{\tab}[1]{Table\,\ref{#1}}
\newcommand{\sct}[1]{Section\,\ref{#1}}

\newcommand{\ctg}{\ensuremath{c_{tg}}}

\newcommand{\pt}{\ensuremath{p_T}\xspace}
\newcommand{\ptH}{\ensuremath{p^H_T}\xspace}
\newcommand{\ptV}{\ensuremath{p^V_T}\xspace}

\newcommand{\lsim}{\raisebox{-0.13cm}{~\shortstack{$<$ \\[-0.07cm]
      $\sim$}}~}

\usepackage{array}
\newcolumntype{L}[1]{>{\raggedright\let\newline\\\arraybackslash\hspace{0pt}}m{#1}}
\newcolumntype{C}[1]{>{\centering\let\newline\\\arraybackslash\hspace{0pt}}m{#1}}
\newcolumntype{R}[1]{>{\raggedleft\let\newline\\\arraybackslash\hspace{0pt}}m{#1}}

\begin{document} 
\begin{titlepage}
\begin{flushright}
MPP-2021-135\\
PSI-PR-21-18\\
ZU-TH 37/21
\end{flushright}

\renewcommand{\thefootnote}{\fnsymbol{footnote}}

\begin{center}
  {\Large \bf Sensitivity to BSM effects\\[0.5cm] in the Higgs \boldmath{$p_T$} spectrum within SMEFT}
\end{center}

\par \vspace{2mm}
\begin{center}
  {\bf Marco Battaglia${}^{(a)}$, Massimiliano Grazzini${}^{(b)}$,\\[0.3cm]
Michael Spira${}^{(c)}$}
and    
{\bf Marius Wiesemann${}^{(d)}$}

\vspace{3mm}

${}^{(a)}$Santa Cruz Institute for Particle Physics, UCSC, Santa Cruz, CA 95064, USA\\[0.25cm]

${}^{(b)}$Physik Institut, Universit\"at Z\"urich, CH-8057 Z\"urich, Switzerland\\[0.25cm]

${}^{(c)}$Paul Scherrer Institut, CH-5232 Villigen PSI, Switzerland\\[0.25cm]

$^{(d)}$Max-Planck-Institut f\"ur Physik, 80805 M\"unchen, DE-80805 Germany

\vspace{5mm}

\end{center}

\par \vspace{2mm}
\begin{center} {\large \bf Abstract}

\end{center}
  
\begin{quote}
  \pretolerance 10000

  The study of Higgs boson production at large transverse momentum is one of the new frontiers for the LHC Higgs physics programme. This paper considers boosted Higgs production in the Standard Model Effective Field Theory (SMEFT). We focus on the gluon fusion and $t{\bar t}H$ production processes and study the effects of three dimension-6 operators: the top Yukawa operator, the gluon-Higgs effective coupling and the chromomagnetic dipole operator of the top quark. We perform a detailed study of the sensitivity of current and future LHC data to the corresponding Wilson coefficients, consistently accounting for their renormalisation group evolution. We compare the sensitivities obtained with only linear and linear + quadratic terms in the SMEFT by using the spectrum shape and the addition of the Higgs signal yields. We also consider fits of $p_T$ spectra in models with heavy-top partners and in MSSM scenarios with a light scalar top and study 
the validity of the SMEFT assumptions as a function of the new-particle masses and the Higgs $p_T$ range. Finally, we extract constraints on the Wilson coefficients for gluon fusion from a simultaneous fit to the ATLAS and CMS data and compare our results with those obtained in global SMEFT analyses.

\end{quote}

\vspace*{\fill}
\begin{flushleft}

September 2021
\end{flushleft}
\end{titlepage}

\section{Introduction}
\label{sec:intro}

After the discovery of the scalar resonance with a mass of 125\,GeV~\cite{ATLASdisc, CMSdisc} the study of its properties has been one of the main activities of the LHC physics programme.
The measurements carried out up to now have shown that, within the current accuracy, the new resonance has properties compatible with those predicted for the Higgs boson in the Standard Model (SM)
and no deviations from the SM picture emerged so far.

At the theoretical level there are two major options to explore scenarios beyond the SM (BSM): either by introducing extensions of the SM particle content through explicit BSM models, or by parametrising deviations from the SM picture supplementing the Lagrangian with higher-dimensional operators built from SM fields.
With the assumption that new physics fulfils the decoupling theorem \cite{Appelquist:1974tg}, the effect of these operators is suppressed by powers of the new-physics scale $\Lambda$~\cite{Burges:1983zg, Leung:1984ni, Buchmuller:1985jz,Grzadkowski:2010es}. The scale $\Lambda$ might be a heavy new-particle mass or a combination thereof, if there are several new particles beyond the SM.
This approach is model independent and, when combined with the SM gauge symmetries, it defines the SM Effective Field Theory (SMEFT) (see \citere{Brivio:2017vri} for a review).\footnote{An even more general approach is the one in which the $SU(2)_L\otimes U(1)$ gauge symmetry is assumed to be realised non linearly and the Higgs field is introduced as scalar singlet. This is the so called Higgs EFT (HEFT) (see e.g. \cite{Feruglio:1992wf,Brivio:2013pma,Alonso:2015fsp,Buchalla:2015qju,Helset:2020yio,Cohen:2020xca}), and can be used to deal with non-decoupling BSM scenarios.}

%
%
%
%
%
%
%
%

A number of SMEFT studies have been carried out in the Higgs sector in recent years, and a considerable amount of work has been devoted to develop appropriate tools to include dimension-six operators \cite{Alloul:2013naa,Artoisenet:2013puc,Contino:2014aaa,Brivio:2017btx,deBlas:2019okz,Degrande:2020evl}. SMEFT analyses including a broad range of measurements, in the Higgs sector and beyond, have been presented in \citeres{Corbett:2012ja,SILH,Pomarol:2013zra,Falkowski:2014tna,Ellis:2014jta,Ellis:2018gqa,Ellis:2020unq,Ethier:2021bye}.

The amount of data collected at the LHC in Run\,2 and the perspectives for Run\,3 and the High Luminosity programme (HL-LHC) offer the possibility to study the {\it dynamical} properties of the Higgs boson in different kinematic regimes.
This is achieved by analyzing Higgs processes involving a large momentum scale inside the relevant matrix element.
Of particular interest is the region in which the Higgs boson is produced at large transverse momentum.
Measurements in this region shed light on the structure of the interactions of the Higgs boson with strongly interacting particles and might unveil BSM effects that are not revealed
through inclusive measurements.

Higgs boson production at large transverse momentum resolves the structure of the Higgs coupling to gluons and possible BSM effects are modeled within SMEFT through effective operators that modify the shape of the spectrum.
The SMEFT approach does not specify the UV completion and has a finite range of validity in the Higgs transverse momentum (\ptH)
which depends on the scale $\Lambda$.
On the other hand, if the experimental accuracy at large \ptH is not sufficient, the measurements will not be sensitive to effects originating from the detailed nature of the UV-completion. Conversely, the theoretical accuracy of the SM cross section will also limit the BSM sensitivity of these measurements. The (universal) interpretation in terms of SMEFT Wilson coefficients will in addition be limited by the theoretical precision of SMEFT effects beyond the SM, and, therefore, the consistent inclusion of higher-order corrections to SMEFT contributions is also relevant \cite{Passarino:2012cb}.

The total Higgs-production cross section in the dominant production channel, namely gluon fusion ($ggF$), is not sufficient to resolve all combinations of the dimension-6 Wilson coefficients. Only with the inclusion of additional production modes, such as the associated production with a top-antitop pair ($t{\bar t}H$), and by considering more exclusive observables, such as differential cross sections, it is possible to disentangle the effects of the relevant dimension-6 operators. The Higgs transverse-momentum distribution and its shape are particularly sensitive to different combinations of the Wilson coefficients at large transverse momentum.

The ATLAS and CMS collaborations have already reported measurements of the Higgs transverse-momentum spectrum~\cite{Aaboud:2018ezd,Sirunyan:2018sgc,Aaboud:2018xdt,ATLAS:2020wny} from the analysis of the full LHC Run\,2 data. They have also presented dedicated boosted $H \rightarrow b \bar b$ analyses aiming to access Higgs production at large transverse momentum~\cite{Sirunyan:2020hwz,Aad:2019uoz,ATLAS-CONF-2021-010}. This provides us with a first set of experimental data to analyse and renders it possible to obtain rather accurate projections of the experimental uncertainties on the Higgs transverse-momentum spectrum in measurements throughout the LHC programme up to the HL-LHC~\cite{CMS:2018qgz,ATL-PHYS-PUB-2018-040}.

The leading order (LO) transverse-momentum distribution of the Higgs boson in $ggF$ production is known since long time including the full quark-mass dependence \cite{ptLO1,ptLO2}. The NLO QCD corrections have first been obtained in the heavy-top-limit (HTL) and have the effect of increasing the differential cross section by roughly a factor of two~\cite{ptNLO0,ptNLO1,ptNLO2,ptNLO3}. At NLO, finite top-mass effects have been estimated in terms of a large top-mass expansion \cite{Harlander:2012hf, Neumann:2014nha}. They have later been supplemented by the inclusion of the full top-mass dependence in the real corrections \cite{Neumann:2016dny,Frederix:2016cnl}. Further improvements of the virtual contribution were obtained in \citeres{Lindert:2018iug,Neumann:2018bsx} through the use of an asymptotic expansion. The exact NLO QCD calculation has eventually been completed in \citere{Jones:2018hbb} through the numerical integration of the corresponding two-loop diagrams.
The next-to-next-to-leading order (NNLO) QCD corrections have been determined in the HTL leading to a moderate increase of the differential cross section and a significant reduction of the residual scale dependence~\cite{Boughezal:2015dra,Boughezal:2015aha,Chen:2016zka}.
The NLO and approximate NNLO theoretical predictions at high-$p_T$ have been combined in \citere{Becker:2020rjp} and compared with the predictions from the other production channels and with the predictions of commonly used event generators \cite{Alioli:2008tz,Campbell:2012am,Hamilton:2012rf,Frederix:2016cnl}.
The transverse momentum spectrum involving additional dimension-6 (and dimension-8) operators has been studied in \citeres{Azatov:2013xha,Grojean:2013nya,Langenegger:2015lra,Maltoni:2016yxb, Deutschmann:2017qum, Grazzini:2016paz, Grazzini:2018eyk,ptdim81,ptdim82}.

In this paper Higgs boson production at large \ptH{} is studied in the framework of SMEFT.
Neglecting the bottom- and charm-loop contributions, which are irrelevant at large transverse momentum, we focus on three operators:
a point-like coupling of the Higgs boson to gluons, a modification of the top-quark Yukawa coupling and the chromomagnetic dipole operator, that modifies the coupling between gluons and the top quark, with and without the Higgs boson at the same vertex.
As far as the gluon fusion channel is concerned, the calculation including the effects of these operators has been performed in \citeres{Grazzini:2016paz,Deutschmann:2017qum,Grazzini:2018eyk}. Besides gluon fusion, the $t{\bar t}H$ channel is also sensitive to the same operators and, at large $p_T$, gives a non-negligible contribution in the experimental analyses. Therefore, we have included this production mode in our study, by carrying out the corresponding calculation in SMEFT. For the other production modes, that are not affected by the above operators, we just stick to the SM predictions.

The central goal of the present study is to place bounds on the relevant 
Wilson coefficients of the dimension-6 operators from current LHC data and to
assess the sensitivity on those Wilson coefficients that can be reached in the 
remainder of the LHC programme. 
To this end, we exploit a calculation of the $p_T$ spectrum in SMEFT in 
combination with available 
state-of-the-art SM predictions \cite{Becker:2020rjp} for all the 
relevant channels, and we perform multi-parameter fits of the deviations from the SM 
transverse-momentum spectrum modelled by the Wilson coefficients that can be 
resolved within the experimental and theoretical uncertainties. In this context, we also 
compare our results with those obtained with analyses in the top sector 
\cite{Brivio:2019ius}, in particular for the chromomagnetic dipole operator. 
Constraints from the present results are obtained with a simultaneous EFT fit to the 
preliminary ATLAS and the CMS data.

In our analysis we probe scales from the Higgs mass up to $1$ TeV, and it is therefore
important to consistently include the renormalisation group evolution of the Wilson coefficients.
This is done by solving the corresponding RG equations \cite{Jenkins:2013zja,Jenkins:2013wua,Alonso:2013hga} at leading-logarithmic (LL) accuracy.

Since the experimental sensitivity is approaching \ptH values of order of 1\,TeV, it becomes mandatory to assess the validity range of the EFT fits 
at the upper end of the Higgs transverse-momentum spectrum when 
extracting information on explicit models. We perform such study
by fitting the \ptH spectra in a new-physics model with a heavy top partner and in the MSSM with a light scalar top through a linear and a non-linear EFT expansion of the dimension-6 operators. We then compare,
as a function of the upper value of the \ptH fit range, 
the values of the Wilson coefficients extracted from our fits to those obtained by
matching the explicit models to the EFT when integrating out the heavy degrees of freedom.

The paper is organised as follows. In \sct{sec:theo} we introduce the theoretical framework and the calculations on which our analysis is based. \sct{sec:fit} describes the methodology of the EFT fits, and \sct{sec:range} discusses the range of validity of the EFTs. In \sct{sec:results} we report the extraction of the Wilson coefficients, we assess the sensitivity that can be reached through the LHC programme and the constraints that can be derived with the current LHC data. \sct{sec:concl} gives our conclusions. The appendix is devoted to the discussion and solution of the renormalisation group equations for the Wilson coefficients.
 
\section{Theoretical Framework}
\label{sec:theo}

In the SMEFT the full set of 2499 dimension-6 operators built from SM fields is added to the SM Lagrangian:
\begin{equation}
{\cal L}_{\rm eff} = {\cal L}_{\rm SM} + \sum_i \frac{c_i}{\Lambda^2} {\cal O}_i\,,
\end{equation}
where $\Lambda$ denotes the scale of new physics, ${\cal O}_i$ are the 
dimension-6 operators and $c_i$ the corresponding Wilson coefficients.
Neglecting the bottom- and charm-loop contributions\footnote{Modifications of the bottom and charm Yukawa coupling have a significant impact only at small transverse momenta of the Higgs boson in gluon fusion, while the bottom Yukawa can be accessed also via the $b\bar{b}H$ production mode \cite{Dittmaier:2003ej, Dawson:2003kb, Harlander:2011fx, Harlander:2010cz,Wiesemann:2014ioa,Harlander:2014hya,Deutschmann:2018avk}, and through the decay $H \rightarrow b\bar{b}$.}, which have no impact at large \pt{}, the following three operators are relevant for our study of the transverse-momentum distribution of the Higgs boson in gluon fusion:
\begin{eqnarray}
{\cal O}_1 & = & |H|^2 G^a_{\mu\nu}G^{a,\mu\nu}\,, \nonumber \\
{\cal O}_2 & = & |H|^2 \bar{Q}_L H^c t_R + h.c.\,, \\ \nonumber
{\cal O}_3 & = & \bar{Q}_L H \sigma^{\mu\nu}T^a t_R G_{\mu\nu}^a + h.c.\,,
\end{eqnarray}
which, for single-Higgs production, can be expanded as:
\begin{eqnarray}
\displaystyle
\frac{c_1}{\Lambda^2}\,{\cal O}_1 & \rightarrow & \frac{\as}{\pi v} c_g h
G^a_{\mu\nu}G^{a,\mu\nu}\,, \nonumber \\
\displaystyle
\frac{c_2}{\Lambda^2}\,{\cal O}_2 & \rightarrow & \frac{m_t}{v} (1-c_{t}) h
\bar{t} t\,,  \label{eq:OPs}\\\nonumber
\displaystyle
\frac{c_3}{\Lambda^2}\,{\cal O}_3 & \rightarrow & c_{tg}\frac{g_S m_t}{2v^3}
(v+h)G_{\mu\nu}^a({\bar t}_L\sigma^{\mu\nu}T^a t_R+h.c)\,,
\end{eqnarray}
with $\sigma^{\mu\nu} = \frac{i}{2} [\gamma^\mu,\gamma^\nu]$ and
$G^a_{\mu\nu} = \partial_\nu G^a_\mu - \partial_\mu G^a_\nu +g_s f_{abc}
G^b_\mu G^c_\nu$. The operator ${\cal O}_1$ describes a point-like
contact interaction between the Higgs boson and gluons, the operator
${\cal O}_2$ corresponds to a modification of the top Yukawa coupling
and the operator ${\cal O}_3$ denotes the chromomagnetic dipole operator
that modifies the coupling between gluons and the top quark, with and
without the Higgs boson at the same vertex.


At large transverse momenta, the \ptH spectrum receives contributions from several production modes.
For the $ggF$ contribution we employ our calculation of the \ptH spectrum presented in \citere{Grazzini:2016paz},
which we have implemented in the numerical program {\sc MoRe-HqT}. With {\sc More-HqT}
the transverse-momentum spectrum at small \pt{} is evaluated at next-to-leading-logarithmic (NLL) accuracy and matched to the LO result at high \pt{}, while being NLO accurate for the cumulative cross section in \pt{}, so as to achieve NLL+NLO accuracy. The program {\sc MoRe-HqT} is based on 
the {\sc MoRe-SusHi} code \cite{Mantler:2012bj,Harlander:2014uea,Liebler:2016dpn}, which computes the NLL+NLO \pt{} spectrum in the SM with full quark-mass dependence 
\cite{Mantler:2012bj} and in models with extended 
Higgs sectors, in particular a simple 2-Higgs Doublet Model, in the MSSM \cite{Harlander:2014uea}, or in the NMSSM \cite{Liebler:2016dpn}.
The contribution of the chromomagnetic operator $O_3$ has also been  implemented \cite{Grazzini:2018eyk}.
For the resummed contribution at small \ptH  the program relies on the $b$-space implementation 
of the {\sc HqT} code \cite{Bozzi:2005wk,deFlorian:2011xf}.
In the present study we focus on the large-\ptH region,
where the spectrum is effectively described at LO, but with the  
inclusion of all relevant dimension-6 operators.

The Wilson coefficients are scale dependent quantities and they obey
renormalization group equations (RGEs) \cite{Jenkins:2013zja,Jenkins:2013wua, Alonso:2013hga}.
We have solved such RGEs at LL accuracy and consistently included the result in our calculation.
The complete derivation is outlined in Appendix\,\ref{sec:appendix}. The solutions read
\begin{align}
c_t(Q^2) & =  c_t(\mu_0^2) +
\frac{24}{5}~\frac{m_t^2(\mu_0^2)}{v^2}~c_{tg}(\mu_0^2)~\left\{
\left(\frac{\alpha_s(Q^2)}{\alpha_s(\mu_0^2)} \right)^{\frac{5}{6\beta_0}} - 1
\right\}\,,\nonumber \\
c_{tg}(Q^2) & = c_{tg}(\mu_0^2) \left(\frac{\alpha_s(Q^2)}{\alpha_s(\mu_0^2)}
\right)^{-\frac{7}{6\beta_0}}\,,
\label{eq:running}\\
c_g(Q^2) & =  \frac{\beta_0 + \beta_1 \alpha_s(Q^2)/\pi}{\beta_0 + \beta_1
\alpha_s(\mu_0^2)/\pi} \left\{ c_g(\mu_0^2) -
\frac{3\pi}{5-6\beta_0}~\frac{m_t^2(\mu_0^2)}{v^2}~
\frac{c_{tg}(\mu_0^2)}{\alpha_s(\mu_0^2)}~\left[
\left(\frac{\alpha_s(Q^2)}{\alpha_s(\mu_0^2)} \right)^{
  \frac{5}{6\beta_0}-1} - 1 \right] \right\}\,,\nonumber
\end{align}
where $\mu_0$ is the input scale at which the Wilson coefficients are extracted and
$Q$ is the actual dynamical scale at which the operators are evaluated
for a given process, that we identify with the 
renormalization scale $\mu_R$ of the strong coupling constant, $\alpha_s$, 
of the process.
We evaluate \eqn{eq:running} following the same strategy as outlined in 
\citere{LHCHiggsCrossSectionWorkingGroup:2016ypw}. In particular, that means that the $\msbar$ mass of the top quark at the scale $\mu_0$,
$m_t(\mu_0^2)$, is evaluated 
by converting the on-shell top mass to $m_t(m_t)$ with the 4-loop expression
of \citere{Marquard:2016dcn} and then evolving with 4-loop running to $m_t(\mu_0^2)$ \cite{Chetyrkin:1997dh,Vermaseren:1997fq}. For 
$\alpha_s$, on the other hand, we use the 2-loop running as given by the 
employed set of parton densities.

Since the same effective operators also enter the $t\bar tH$ production mode, which has a non-negligible contribution to the transverse-momentum spectrum of the Higgs boson at large \pt{}, we compute the transverse-momentum spectrum with the inclusion of the \mbox{dimension-6} operators of \eqn{eq:OPs}. We have performed this calculation both
analytically and by using {\sc MadGraph5\_aMC@NLO} \cite{Alwall:2014hca}. For the latter the generation of the relevant matrix elements has been carried out by using as input
the {\tt SMEFTatNLO} Unified FeynRules Output (UFO) model \cite{Degrande:2020evl}.
The {\sc MadGraph5\_aMC@NLO} results for $t{\bar t}H$ production have been compared with our independent analytical calculation finding complete agreement.\footnote{We note that with {\sc MadGraph5\_aMC@NLO} the $t{\bar t}H$ calculation could in principle be carried out at NLO. However, since the computation for the $ggF$ process can only be performed at LO, and, moreover, the $t{\bar t}H$ contribution is subdominant, we limit ourselves to use our LO result.}
The translation of the {\sc MadGraph5\_aMC@NLO} conventions for the Wilson coefficients into our notation is derived through\footnote{The advantage of our convention is explicitly visible in the simplicity of the RGEs derived in Appendix A, i.e.~the leading effects of the QCD running are factorized in terms of the $\alpha_s$ and $m_t$ factors that lead to a pure BSM evolution of our coefficients $c_t, c_g, c_{tg}$ at LL level, if $c_{tg}$ does not vanish at the input scale.}
\begin{align}
{\tt cpG} & =  \frac{\alpha_s}{\pi}~\frac{\Lambda^2}{v^2}~c_g \,, \nonumber \\
{\tt ctp} & =  \frac{\sqrt{2} m_t}{v}~\frac{\Lambda^2}{v^2}~
(1-c_t) \,,  \label{eq:madconv}\\\nonumber
{\tt ctG} & =  \frac{m_t}{\sqrt{2} v}~\frac{\Lambda^2}{v^2}~c_{tg} \,.
\end{align}
Since we set the scale of the operators to be equal to the renormalization scale $\mu_R$, this translation has
to be carried out at the scale $\mu_R$ used in the $t{\bar t}H$ calculation. Moreover, given that the factor $\sqrt{2}\,m_t/v$ originates from the 
Yukawa coupling, whose renormalization is on-shell in the $t{\bar t}H$ calculation, we set $m_t$ to the value of the pole mass in the conversion of \eqn{eq:madconv}. This ensures that, disregarding running effects, $c_t$ remains a rescaling of the SM Yukawa coupling.
Finally, we include the running of the Wilson coefficients also in the $t\bar{t}H$ calculation by evolving the operators from the input scale $\mu_0$ to $\mu_R$ through \eqn{eq:running} before converting to the {\sc MadGraph5\_aMC@NLO} conventions with \eqn{eq:madconv}.

The chromomagnetic dipole operator is of particular interest because it can be tested also in the top sector \cite{Franzosi:2015osa,Brivio:2019ius}.
The corresponding Wilson coefficient $C_{tG}$ of these works is related to our $c_{tg}$ as
\begin{equation}
\label{eq:convctg}
  \frac{C_{tG}}{\Lambda^2}=c_{tg}\frac{g_s\, m_t}{\sqrt{2}v^3}\, .
\end{equation}
In \eqn{eq:convctg} the scale of $g_s=\sqrt{4\pi\alpha_s}$, of the $\msbar$ top mass $m_t$ and of the Wilson coefficients should be understood as $m_t$.
To the best of our understanding, in the analyses of \cite{Franzosi:2015osa,Brivio:2019ius} the running of the Wilson coefficients is neglected and the scale $\Lambda$ is set to 1 TeV. 
In this work we include the effect of the QCD running as discussed above.
When the transverse momenta span over a large 
range (for instance for the \ptH spectrum from ${\cal O}(100)$ GeV to $1-2$ TeV),
it is essential to include the QCD running of the Wilson coefficients in order to
consistently perform an EFT analysis.

Our computations of the high-\ptH cross section in gluon fusion and in $t{\bar t}H$ production are used to obtain the BSM effects as ratios of the Higgs cross section with SMEFT corrections, $\sigma_{i}(c_t,c_g,c_{tg})$, to the SM prediction, $\sigma_{i}^{\rm SM}$:
\begin{align}
  \label{eq:ratio}
  R_i (c_g,c_{tg},c_t)= \Delta \sigma_{i}(c_t,c_g,c_{tg})/ \Delta \sigma_{i}^{\rm SM}\,,
\end{align}
 where $\Delta \sigma_{i}$ is the differential cross section integrated in each \ptH bin $i$ and both numerator and denominator are computed at the same order in perturbation theory.
 These ratios, computed on a grid of values of $c_g$, $c_{tg}$ and $c_t$, are employed
 to study the potential of Higgs measurements at large \ptH to constrain the effects of the dimension-6 operators.

We specify the parameters used for all predictions obtained in the remainder of this paper.
We use the five-flavour scheme with the corresponding NLO set of the PDF4LHC2015 \cite{Butterworth:2015oua,Ball:2014uwa,Dulat:2015mca,Harland-Lang:2014zoa,Gao:2013bia,Carrazza:2015aoa} parton distribution functions~(PDFs) and the respective value of the strong coupling constant. 
The top-quark pole mass is set to $172.5$\,GeV.
Since the $R_i (c_g,c_{tg},c_t)$ ratios are computed in the LO approximation, they are independent of the overall powers of the QCD coupling $\as$
and their respective setting of the renormalisation scale $\mu_R$. We also 
find that they are largely independent of the choice of the factorisation scale $\mu_F$, since the parton densities approximately cancel in the ratio, 
and they also show only a mild dependence on the scale at which the Wilson coefficients are evaluated, which we also choose as $\mu_R$. We evaluate the ratios $R_i$ in \eqn{eq:ratio} by dynamically setting $\mu_R=\mu_F=m_T$, where $m_T$ denotes the transverse mass of the Higgs boson $m_T = \sqrt{m_H^2+p_{TH}^2}$, for the $ggF$ process. In the case of $t\bar{t}H$ production the scales are set to $\mu_R=\mu_F=(m_H+2\,m_t)/2$.

We stress that the choice of a dynamic scale together with Eq.~(\ref{eq:running}) implies that even though $c_g$ and $c_t-1$ may vanish at the input scale $\mu_0$, a non-zero 
value is generated by $c_{tg}$ through their running to the scale 
$Q$ that depends on the bin in \ptH under consideration.
 As a result, $c_t$ is not a simple rescaling of the Higgs cross 
section, but its running renders it dependent on \ptH and therefore it will also affect the shape of the Higgs transverse-momentum
spectrum, although the effect should be subleading.
This also implies that in every given bin in \ptH, the values of
$c_g$, $c_t$ and $c_{tg}$ differ, and that what we extract in our fits
are the Wilson coefficients at the input scale, 
namely $c_g(\mu_0)$, $c_t(\mu_0)$ and $c_{tg}(\mu_0)$. In order to 
obtain the Wilson coefficients at any other scale we may again employ 
their evolution in \eqn{eq:running}. This is completely analogous to 
the treatment of the strong coupling constant $\alpha_s$ 
in its measurements or to simultaneous
extractions of $\alpha_s$ and the top-quark mass $m_t$, where their
QCD running is a crucial consistency aspect of the respective template fits.
We also stress that our definition of the dimension-6 operators in 
\eqn{eq:OPs} is particularly convenient, as it expresses the running of 
the operators by the one of $\alpha_s$ and $m_t$, and limits the 
mixing terms to contributions induced by $c_{tg}$.

\section[Fitting the Higgs $p_T$ Spectrum]{Fitting the Higgs \boldmath{$p_T$} Spectrum}
\label{sec:fit}
The Wilson coefficients $c_g$, $c_{tg}$ and $c_t$ are extracted from the Higgs \pt spectrum by using a multi-parameter fit of the predicted deviations $R_i(c_g, c_{tg}, c_t)$ for each bin $i$ in \ptH to the ratio of the measured cross section over the best
SM prediction in each bin, while taking into account the experimental and theoretical uncertainties. To compute $R_i(c_g, c_{tg}, c_t)$ we have generated the
Higgs transverse-momentum spectra for $ggF$ production with {\sc MoRe-HqT} on a three dimensional grid for $-1.0\le c_g \le 1.0$, $-0.5\le c_{tg} \le 0.5$, and $0.1\le c_t \le 2.0$, and on bins of 50~GeV in the Higgs transverse momentum. 
The range in $c_{g}$ and $c_{tg}$ is chosen according to the sensitivity of the current dataset.
Similarly, the \ptH{} spectra for $t{\bar t}H$ production have been generated with {\sc MadGraph5\_aMC@NLO}, after translating the Wilson coefficients to the conventions used in our study at the scale $\mu_R$ and including their running from $\mu_0$ to $\mu_R$, see \eqn{eq:madconv} and \eqn{eq:running}, respectively. In order to match the binning of the experimental measurements the \ptH{} spectra are rebinned by integrating the cross section values for the chosen bin interval and normalising to the bin width. After rebinning,
the values $R_i(c_g, c_{tg}, c_t)$ are calculated separately for each
production mode ($ggF$ and/or $t{\bar t}H$) by taking
the ratio to the corresponding SM cross sections evaluated with the same tools (i.e.\ at the same order in perturbation theory). The maps, which are defined as the EFT spectra for all values of the Wilson coefficients, are stored for each production process as sets of {\sc Root}~\cite{Brun:1997pa} histograms. 

The fitting code is based on the {\sc Minuit} package~\cite{James:1975dr} in the {\sc Root} framework. Histograms in the map of the EFT spectra generated for a specific scan are loaded in the fitting program as vectors of histograms, whose index is related to the $(c_g, c_{tg}, c_t)$ values used to generate them. Each map of the EFT spectra has a reference SM spectrum used to compute the $R_i$ values.
The spectrum that is fitted can be chosen to be one of the histograms stored in the EFT scan maps, as a histogram from a benchmark model or as the signal strength values reported by the LHC experiments.

\begin{figure}[ht!]
  \begin{center}
    \includegraphics[width=0.65\textwidth]{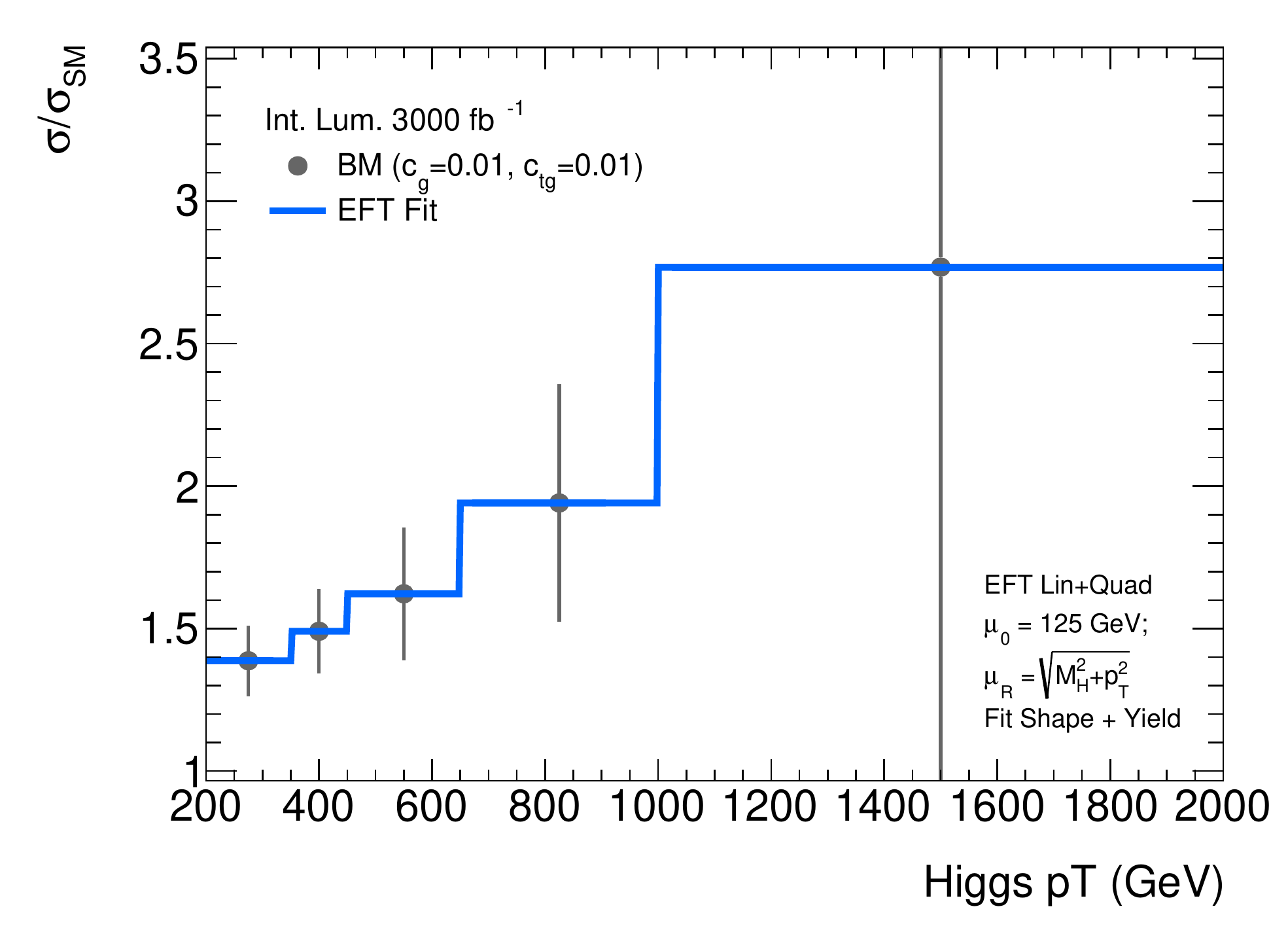} \\ 
    \caption{\label{fig:fitbin} Example of an EFT fit on a benchmark \ptH spectrum (pseudo data) with the binning adopted by the experiments. The $R_i$ deviations from the SM values for the ($c_g(M_H)$ = 0.01, $c_{tg}(M_H)$ = 0.01, $c_t(M_H)$ = 1.0) benchmark spectrum are given by the points with the estimated bin-by-bin uncertainties, which are obtained by combining the predicted ATLAS and CMS accuracies with $12\%$ uncertainty on the SM cross section, while the continuous line shows the result of the EFT fit.}    
    \end{center}
\end{figure}

The fitting procedure minimizes the $\chi^2$ of the fit with respect to the values of the Wilson coefficients. 
Since the grid of the Wilson coefficients used for the EFT scans is discrete, the
$R_i$ values for arbitrary combinations of $(c_g, c_{tg}, c_t)$ within the ranges given above are evaluated in the fitting code by an interpolation between the two closest generated values. When performing fits to the \ptH{} spectrum including the contribution of both $ggF$ and $t{\bar t}H$ production modes the $\chi^2$ is evaluated from the sum of the $ggF$ and $t{\bar t}H$ EFT contributions, each weighted by the assumed fractions $f_{ggF}$, $f_{t{\bar t}H}$ of signal events in the sample. What remains is a SM-like spectrum with a fraction $1 - f_{ggF} - f_{t{\bar t}H}$ for the $VH$ and VBF production modes that is not sensitive to the ${\cal O}_1$, ${\cal O}_2$ and ${\cal O}_3$ EFT operators.

The fit has three free parameters: $c_g$ and $c_{tg}$, defined at $\mu_0$=125~GeV unless stated otherwise, and an overall normalisation factor, approximately corresponding to $c_t^2$, that can be fitted to remove the sensitivity to the integrated cross section. Alternatively, this sensitivity can also be removed by normalising the $R_i$ values to the average $R$ computed in a \pt range below the lower edge of the fit region. The fitting procedure has been extensively validated by generating benchmark ratios $R_i$ of the \ptH{} spectra for various combinations of $(c_g, c_{tg}, c_t)$ within the given parameter space normalised to the reference SM spectrum and comparing the ensuing  fitted values of $(c_g, c_{tg})$ to the input used to generate them. An example of a fitted benchmark spectrum is shown in Figure~\ref{fig:fitbin}, which clearly shows that the fit reproduces the benchmark spectrum and the input values $(c_g, c_{tg})$=(0.01, 0.01) are precisely returned by the fit, yielding $c_g$ = 0.01$\pm$0.04 and $c_{tg}$=0.01$\pm$0.10, where the quoted uncertainties are correlated and simply result from the assumed errors on the benchmark spectrum.

\subsection[Comparing EFT Effects in $ggF$ and $t{\bar t}H$ Production]{Comparing EFT Effects in \boldmath{$ggF$} and \boldmath{$t{\bar t}H$} Production}
\label{sec:prod}

The $ggF$ and $t{\bar t}H$ production modes are sensitive to the same dimension-6 EFT operators. Moreover, at high \ptH the relative contribution of the $t{\bar t}H$ process increases and a considerable fraction of $t{\bar t}H$ events are accepted alongside $ggF$ events by the selection criteria applied in the boosted analyses inclusive over the production modes. The ATLAS collaboration reports that $\simeq$ 15\% of the signal events selected at \pt $>$ 650~GeV originate from $t{\bar t}H$ production \cite{ATLAS-CONF-2021-010}. In view of this, it is important to evaluate the EFT effects on the \ptH spectrum for the Wilson coefficients\footnote{The scaling of the $ggF$ and $t\bar{t}H$ cross section with $c_t^2$ is straightforward, since we only include top-quark loops in the $ggF$ cross section.}  $c_g$, $c_{tg}$ for both $ggF$ and $t{\bar t}H$ events. We compare those effects by computing the $R_i$ ratios in 2-dimensional $c_g$, $c_{tg}$ scan for $ggF$ and $t{\bar t}H$ production separately.

\begin{figure}[t]
  \begin{center}
     \begin{subfigure}{0.45\linewidth}
        \centering
        \includegraphics[width=\linewidth]{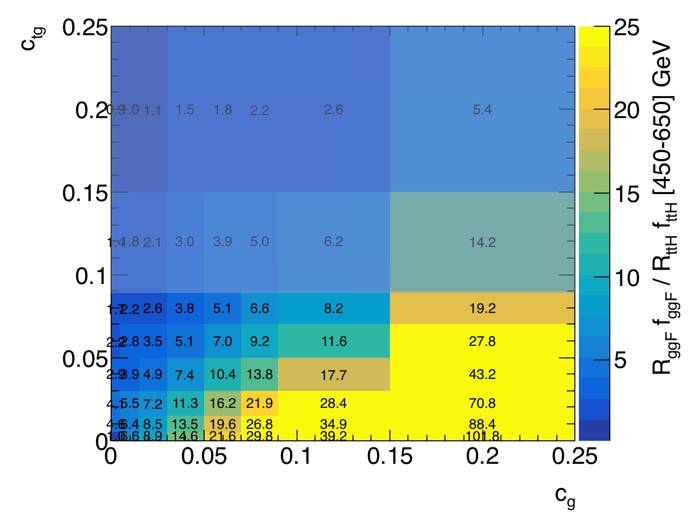}
     \end{subfigure}
     \begin{subfigure}{0.45\linewidth}
        \centering
        \includegraphics[width=\linewidth]{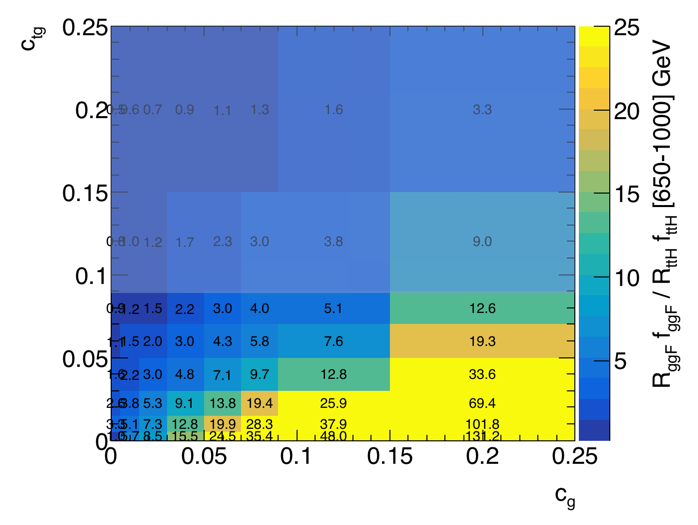}
     \end{subfigure}
   \\[\baselineskip]
     \begin{subfigure}[H]{0.45\linewidth}
       \centering
       \includegraphics[width=\linewidth]{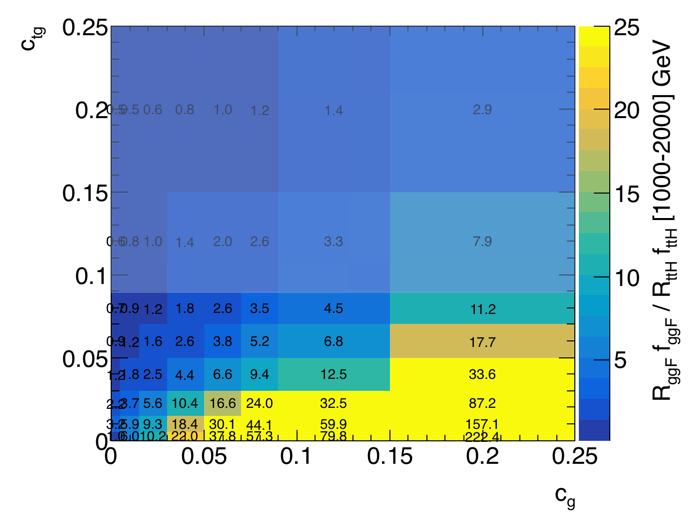}
     \end{subfigure}     
\hfill
    \caption{\label{fig:ggFttH} Ratios of EFT yields normalised to SM predictions in the $ggF$ to $t{\bar t}H$ production processes in the intervals 450 $< p^{H}_T <$ 650~GeV (top left),  650 $< p^{H}_T <$ 1000~GeV (top right) and  1000 $< p^{H}_T <$ 2000~GeV (bottom center), weighted by the signal fractions $f_{ggF}$=0.50 and $f_{t{\bar t}H}$=0.15 of $ggF$ and $t{\bar t}H$ accepted signal events, respectively, corresponding to the values reported for the ATLAS and CMS analyses. The blue shaded area indicates the range of $c_{tg}$ values excluded by the fit of top observables of  Ref.~\cite{Brivio:2019ius} translated to our definition of $c_{tg}(\mu_0)$ for $\mu_0=125$ GeV.}
    \end{center}
\end{figure}

Here we focus on the situation in which both linear and quadratic terms are added in the EFT expansion. When only linear terms are considered the weight of $t{\bar t}H$ production is found to be significantly smaller.
Figure~\ref{fig:ggFttH} shows the ratio of $R_i\times f$ for $ggF$ to the one for $t{\bar t}H$ in three different \ptH bins, where
$f$ is the typical fraction of signal events for the respective production mode accepted in the ATLAS and CMS analyses. Qualitatively, the pattern of enhancement at high \ptH{} is similar for the two production processes. This seems to exclude the possibility of cancellation effects due to opposing contributions from $ggF$ and $t{\bar t}H$. Quantitatively, the effects of the gluon-fusion production process are dominant over the parameter space with the exception of the region with $c_g$ values close to zero, where the enhancement of the $ggF$ process becomes relatively small. In this region, the effects from $ggF$ and $t{\bar t}H$ production are comparable when taking into account the difference in the signal-event fractions $f$, as we do in our analysis.
Large values of $c_{tg}$, which would correspond to a large enhancement of EFT effects in the $t{\bar t}H$ channel, are excluded by precision measurements in the top sector \cite{Brivio:2019ius}, as indicated by the blue-shaded area in the plots. In combination with the larger fractions of $ggF$-produced Higgs bosons selected by the experimental analyses, this bound ensures that $ggF$ dominates the effects that can be detected at the LHC in boosted Higgs production with the current accuracy. However, $t{\bar t}H$ contributions cannot be neglected as the experimental accuracy will improve. A quantitative assessment is presented in \sct{sec:procsens}.

Another aspect that needs to be taken into account in the interpretation of the experimental data are the different shapes of the \ptH spectra for the $ggF$, $t{\bar t}H$, $VH$ and VBF production modes, in particular at their high end of the \ptH{} range.  As an example, a SM \ptH spectrum, consisting of 85\% $ggF$ and 15\% $VH$ production, yields a \ctg value of 0.005, if fitted as pure $ggF$. While this is not yet a concern at the current level of experimental accuracy, the accuracy in the determination of the contribution of the Higgs production processes to the signal sample will become relevant in the study of HL-LHC data, and the $VH$ and VBF production modes should be properly subtracted when
extracting the Wilson coefficients under consideration, as we do in our study.

\section{Validity Range of the SMEFT}
\label{sec:range}
The application of the SMEFT framework assumes that all the New Physics degrees of freedom can be integrated out.
This assumption becomes questionable as the LHC analyses start probing transverse-momentum scales of the order of 1\,TeV, requiring to reconsider the SMEFT validity and the possible implications on explicit BSM models.
The problem of the breakdown of the SMEFT framework when studying Higgs production has been already considered in several studies, e.g\ in \citeres{Grojean:2013nya,PhysRevD.101.115004}. Here, we want to revisit this problem
by considering two explicit models of New Physics impacting the shape of the \ptH spectrum at high values, a heavy top partner model and the MSSM with a light scalar top, and profit from our fitting procedure to draw quantitative conclusions. First, we compute the matching between the explicit models and the SMEFT Wilson coefficients. Both the heavy top partner model and the MSSM with a light scalar top modify only the $c_g$ and $c_t$ Wilson coefficients and do not generate $c_{tg}$ terms at leading order. This also implies that $c_g$ and $c_t$ do not run at LL-level, while $c_g$ does include NLL running through $\as$. Then, we study the shape of the \ptH{} spectrum in these models, i.e.\ the $R_i$ correction to the SM, while varying the masses of the relevant particles, and compare them to those in the SMEFT for the $c_g$ and $c_t$ values computed from the matching. Finally, we perform SMEFT fits to these \ptH{} spectra and study the fitted value of $c_g$ while changing the upper end of the \ptH{} range used in the fit.  The fits are performed using the same \ptH{} binning adopted for the published experimental studies, assuming a flat relative accuracy of 5\% on all the bins, given the small values of the matched Wilson coefficients. This accuracy, whose value is arbitrary and is not meant to reproduce the experimental and theoretical uncertainties, is chosen to ensure that differences of the results obtained while changing the fit range are significant. The comparison of the $c_g$ values from these fits to those obtained via the matching calculation gives quantitative information on the breakdown of the SMEFT approach as a function of the new particle masses and \ptH{} fit range.

\subsection{Heavy Top Partner}
\label{sec:rangeHT}

Several SM extensions predict vector-like quarks, i.e.\ strongly interacting fermions whose left and right-handed components transform in the same way under the $SU(2)\otimes U(1)$ gauge group. These fermions can have different $SU(2)_L$ quantum numbers: in a singlet $(T)$ or $(B)$, in doublets $(T,B)$, in triplets $(X,T,B)$, but they are in the fundamental representation of the colour $SU(3)$  as the ordinary quarks. The effective Lagrangian describing the interaction of the heavy fermions with SM fields contains a Yukawa interaction between new and SM fermions, which induces a mixing between them whose detailed structure depends on the model \cite{delAguila:2000rc}. 
Limits on vector-like partners of third-generation quarks from LHC searches are in the $1.3-1.4$\,TeV range, depending on the scenario under consideration (see e.g. Ref.~\cite{Aaboud:2018pii}).
The mixing with SM fermions leads to strong constraints from precision EW data \cite{Aguilar-Saavedra:2013qpa}.
In particular, heavy quark partners contribute to the $S$ and $T$ oblique parameters \cite{Peskin:1990zt} through mixing. The mass matrices can be diagonalised by suitable biunitary transformations characterised by angles $\theta_R$ and $\theta_L$ (which are not independent).

In the following we consider a simplified model with a top partner with mass $M_T$. The Yukawa couplings in the top sector can be parametrised in terms of a mixing angle $\theta$ and read
\begin{equation}
y_t=\sqrt{2}\,\frac{m_t}{v}\cos^2\theta~~~~y_T=\sqrt{2}\,\frac{M_T}{v}\sin^2\theta\, .
\end{equation}
In the limit $M_T\to \infty$ the top partner can be integrated out and the model is matched to the SMEFT with the following Wilson coefficients:
\begin{align}
  c_g &= \frac{\sin^2 \theta}{12} \,, \nonumber \\
  c_t &= \cos^2 \theta \,, \nonumber \\
  c_{tg} &= 0.
\end{align}

In the following, we consider four $M_T$ mass values: 500, 700, 1500, 2500\,GeV with $\sin^2 \theta$ mixing of 0.1, chosen close to the current experimental limits.

\begin{figure}[t]
  \begin{center}
    \includegraphics[width=0.65\textwidth]{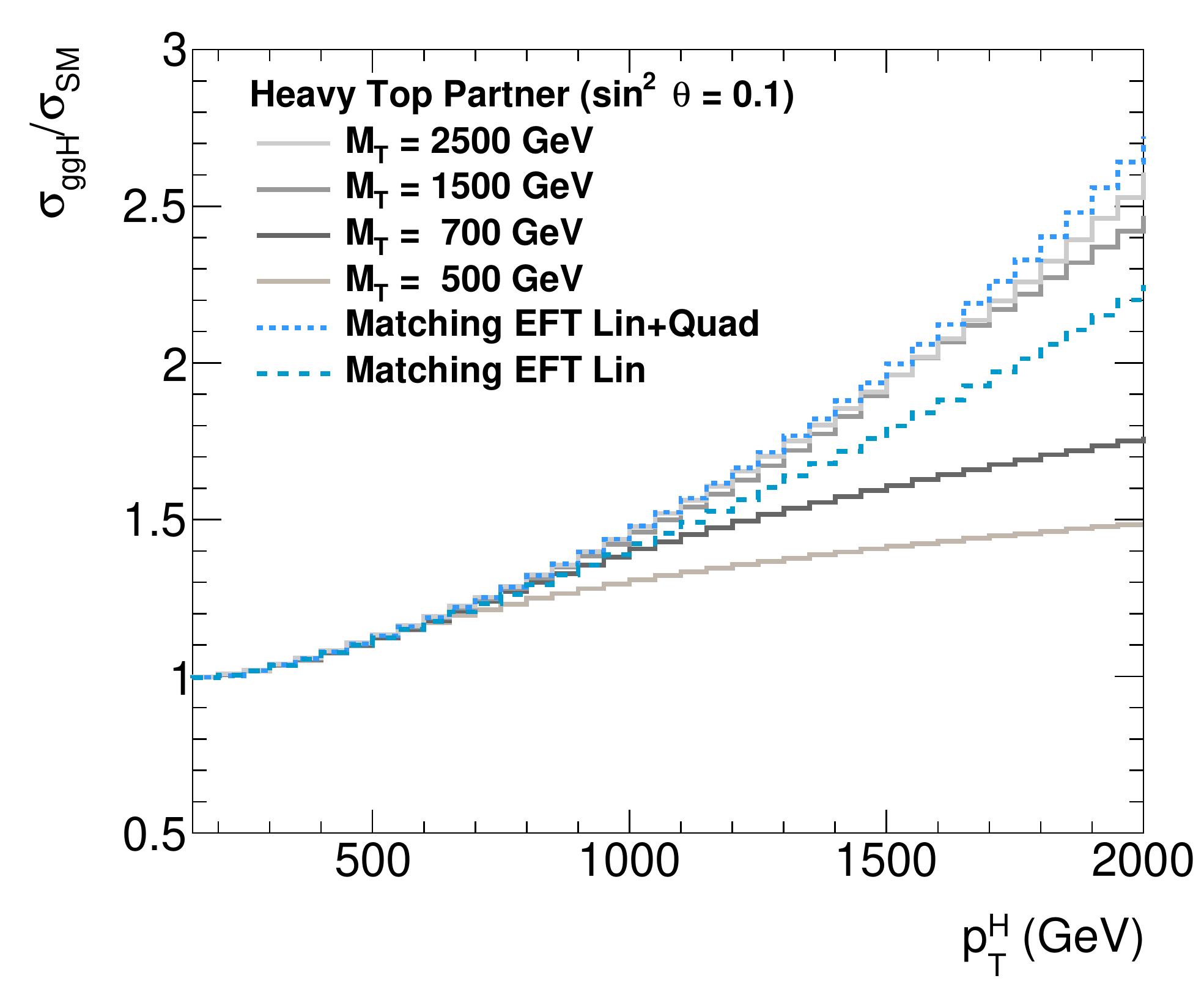}
    \caption{\label{fig:spectraHT} \ptH spectra normalised to the SM prediction for the heavy top partner model with $M_T =$ 500, 700, 1500, 2500\,GeV and $\sin^2 \theta =$ 0.1. The SMEFT spectra for the matched Wilson coefficients and only linear (dashed) and linear plus quadratic (dotted line) terms are also shown.}
  \end{center}
\end{figure}

The \ptH spectra are generated with the {\sc MoRe-HqT} program. The model spectra, normalised to the SM prediction, are shown in \fig{fig:spectraHT}, compared to the SMEFT spectrum for the matched values of $c_g = 0.1/12 \approx 0.0083$ and $c_t = 0.90$. Qualitatively, the matched SMEFT spectrum reproduces that of the model up to \ptH $\lsim M_T$ while at higher \ptH values, where the model spectrum depends explicitly on $M_T^2$ mass terms, the SMEFT description breaks down. The patterns observed here correspond to those reported in Refs.~\cite{Banfi:2013yoa,Grojean:2013nya}.

\begin{figure}[h!]
  \begin{center}
    \begin{tabular}{cc}
      \includegraphics[width=0.475\textwidth]{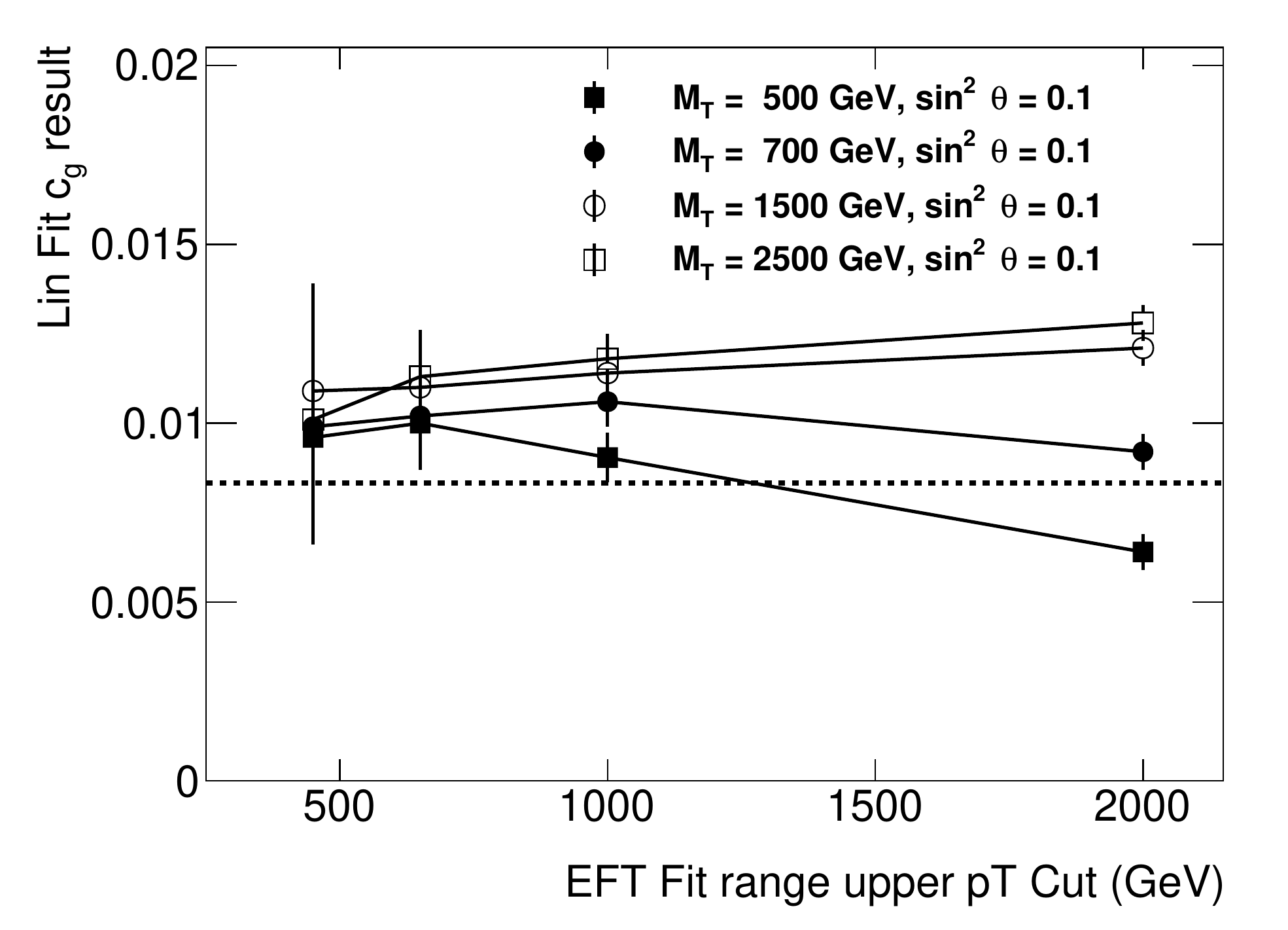} &
      \includegraphics[width=0.475\textwidth]{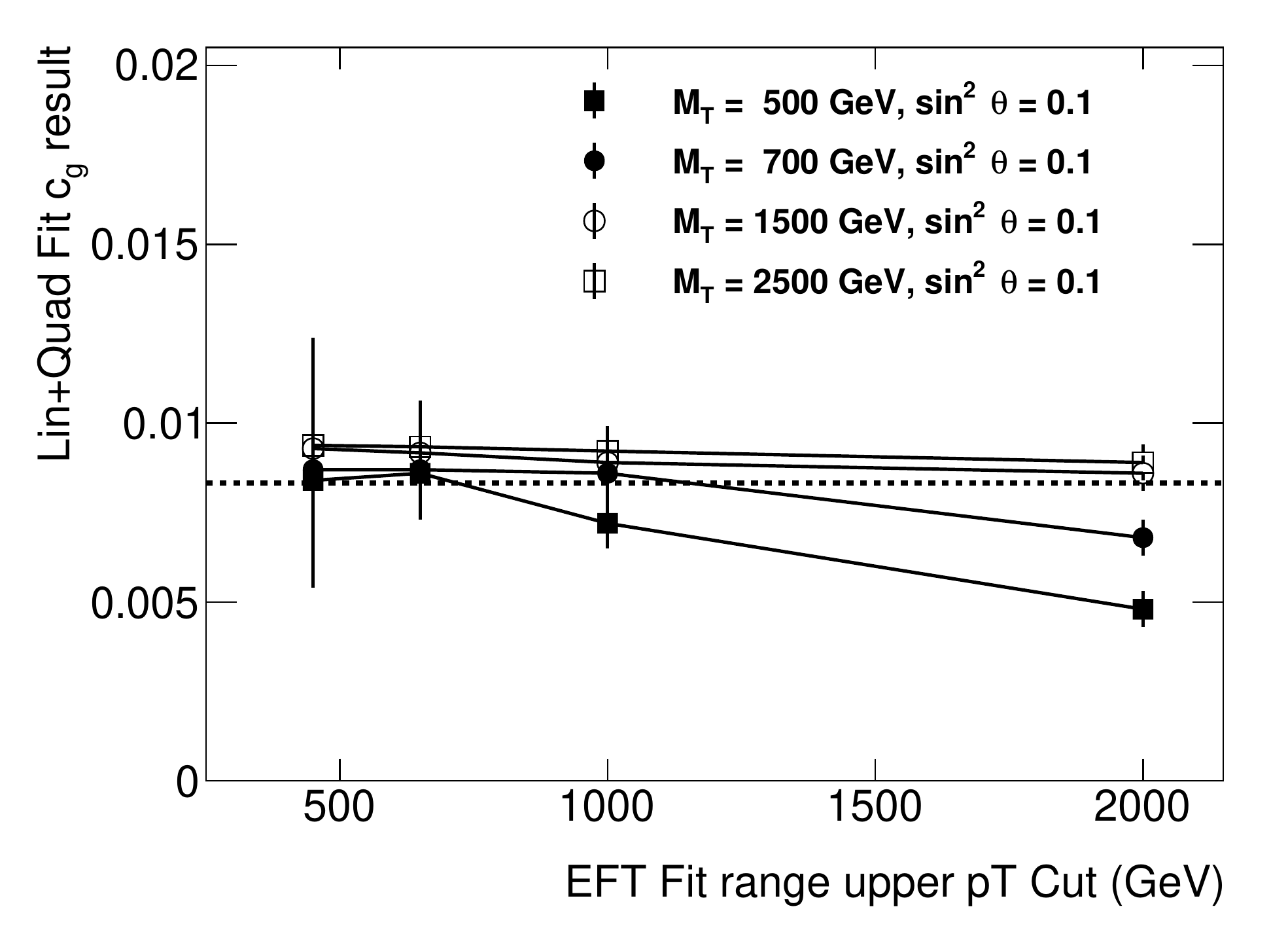} \\
      \end{tabular}
    \caption{\label{fig:fitHT} Wilson coefficient $c_g$ values obtained in the SMEFT fit with only linear (left) and with linear plus quadratic terms (right) to \ptH spectra produced from heavy top partner models with different $M_T$ values and $\sin^2 \theta$=0.1 as a function of the upper end of the \ptH fit range. The dashed lines show the $c_g$ values of the SMEFT matching to the model. }
    \end{center}
\end{figure}

The $c_g$ values corresponding to these spectra are extracted by performing SMEFT fits to the \ptH{} spectra in the heavy top partner model for the chosen values of $M_T$. The fits are performed on a \ptH{} range from 200\,GeV up to an upper limit ranging from 450\,GeV to 2\,TeV. The fitted $c_g$ values are shown in \fig{fig:fitHT} as a function of the upper limit of the fit range for both linear terms and linear plus quadratic terms included in the SMEFT expansion. 
In the latter case, the results show that the $c_g$ values obtained from the SMEFT fits agree with the matched values as long as the upper end of the \ptH range used in the fit is not significantly larger than the heavy top partner mass $M_T$ in the model. When the \ptH range of the fit overlaps with the value of $M_T$ in the model and probes the \ptH region where the SMEFT description breaks down, the fitted $c_g$ values drop below the matched value becoming incompatible with it. This marks the breakdown of the SMEFT to describe the heavy top partner model and it is caused by the mismatch between the spectrum in this model and that from the SMEFT when the fit extends to $\ptH \gtrsim M_T$ values. 
Therefore, the \ptH{} spectrum is not only sensitive to the size of the couplings $c_t, c_g$, but also to the heavy top-partner mass $M_T$, if the accessible \ptH{} values exceed the value of $M_T$.
In the case of the linear SMEFT fit the situation is different. The fitted $c_g$ values have an upward trend with increasing values of the upper bound of the \ptH{} fit range, modulated by the effect of the SMEFT breakdown for $\ptH \gtrsim M_T$, and they do not reproduce the expected value obtained from the matching calculation.

This result is in line with the different behavior of the spectra obtained by matching with only linear or linear plus quadratic terms in \fig{fig:spectraHT}.
In conclusion, the SMEFT expansion with linear and quadratic terms leads to more consistent constraints on the explicit model, as long the $\ptH$ range is sufficiently small compared to $M_T$, than when only the linear terms are kept 
in the SMEFT expansion. This is a clear argument for including quadratic terms in SMEFT fits, which corresponds to an SMEFT expansion at the level of the amplitudes.
On the other hand, the question arises if, in that case, dimension-8 operators should not be included as well, the effects of which are of the same order within the SMEFT expansion at the level of the differential cross section. However, dimension-8 operators decouple with the heavy top-partner mass explicitly, since they constitute the first subleading order of an expansion in the inverse of the heavy top-partner mass at the amplitude level. Thus, the quadratic terms of the dimension-6 operators are dominant.

\subsection{MSSM Scenario with a Light Scalar Top}
\label{sec:rangeMSSM}

The second model we consider is the MSSM with a light scalar top quark (stop), $\tilde{t}_{1}$. The sensitivity of the upper end of the \ptH spectrum to the contribution of $\tilde{t}_{1}$, has already been discussed in some detail~\cite{Grojean:2013nya,Banfi:2018}. 
The change of the shape of the \ptH{} spectrum with the $\tilde{t}_{1}$ mass is estimated for MSSM benchmarks where the masses of all of the SUSY particles are set above 2\,TeV, except for the scalar top and scalar bottom (sbottom) quarks. The sbottom gives only a small contribution to the cross section though. We vary $m_{\tilde{t}_{1}}$ from 500\,GeV, close to current limits from direct searches at the LHC~\cite{CMS:2021eha,ATL-PHYS-PUB-2021-019}, up to 1.5\,TeV. The corresponding MSSM parameters are calculated with {\sc SoftSusy 4.1.7}~\cite{Allanach2002}, and the most relevant to our study are summarised in \tab{tab:mssm}. The \ptH spectrum for the MSSM benchmarks is computed using {\sc SusHi 1.7.0}~\cite{Harlander2013} interfaced with the output of {\sc SoftSusy}.

\begin{table}[t]
  \begin{center}
    \begin{tabular}{|l|c|c|c|}
      \hline
      Parameter & Benchmark               & Benchmark                & Benchmark \\
      & $m_{\tilde{t}}$ 500    & $m_{\tilde{t}}$ 700    &  $m_{\tilde{t}}$ 1500     \\
      \hline
      $m_{\tilde{t}_{1}}$ (GeV) & 512 & 712 & 1572 \\
      $m_{\tilde{t}_{2}}$ (GeV) & 2023 & 2033 & 2068 \\
      $m_{\tilde{b}_{1}}$ (GeV) & 1015 & 1031 & 1082 \\
      $m_{\tilde{b}_{2}}$ (GeV) & 2017 & 2027 & 2059 \\
      $\tan \beta $ & 3.9 & 3.9 & 3.9 \\
      $\mu$ (TeV) & $-3.0$ & $-3.0$ & $-3.0$ \\
      $A_t$ (GeV) & 200 & 200 & 200 \\
      $m_{A^0}$ (TeV) & 2.5 & 2.5 & 2.5 \\
      \hline
    \end{tabular}
  \end{center}
    \caption{\label{tab:mssm} Main parameters of the MSSM benchmark scenarios used in the SMEFT validity study.}
\end{table}

The matched Wilson coefficient $c_g$ can be expressed in terms of the MSSM masses and couplings as:
\begin{eqnarray}
\label{eq:MSSMmatchcg}
  c_g = \frac{1}{96} \,\left( \frac{m_{\mathrm{t}}^2}{m_{\tilde{t}_{1}}^2} g_{\tilde{t}_{1}} + \frac{m_{\mathrm{t}}^2}{m_{\tilde{t}_{2}}^2} g_{\tilde{t}_{2}} + \frac{m_{\mathrm{b}}^2}{m_{\tilde{b}_{1}}^2} g_{\tilde{b}_{1}} + \frac{m_{\mathrm{b}}^2}{m_{\tilde{b}_{2}}^2} g_{\tilde{b}_{2}}\right)\,,
  \end{eqnarray}
where $g_{\tilde{t}}$ and $g_{\tilde{b}}$ are the Higgs couplings to the stop and sbottom quarks in the conventions of {\sc SusHi}, while $c_{tg} = 0$, since the MSSM does not generate $c_{tg}$ terms at leading order. Given the difference in the top and bottom quark masses, the value of $c_g$ for our benchmark points is dominated by the stop contribution. The value of the coefficient $c_t$ is about 0.998 for these scenarios, so that we can limit ourselves to consider the effects of $c_g$. Since $c_{tg}=0$, the Wilson coefficients $c_g$, $c_t$ do not run at LL-level.

\begin{figure}[t]
  \begin{center}
    \includegraphics[width=0.65\textwidth]{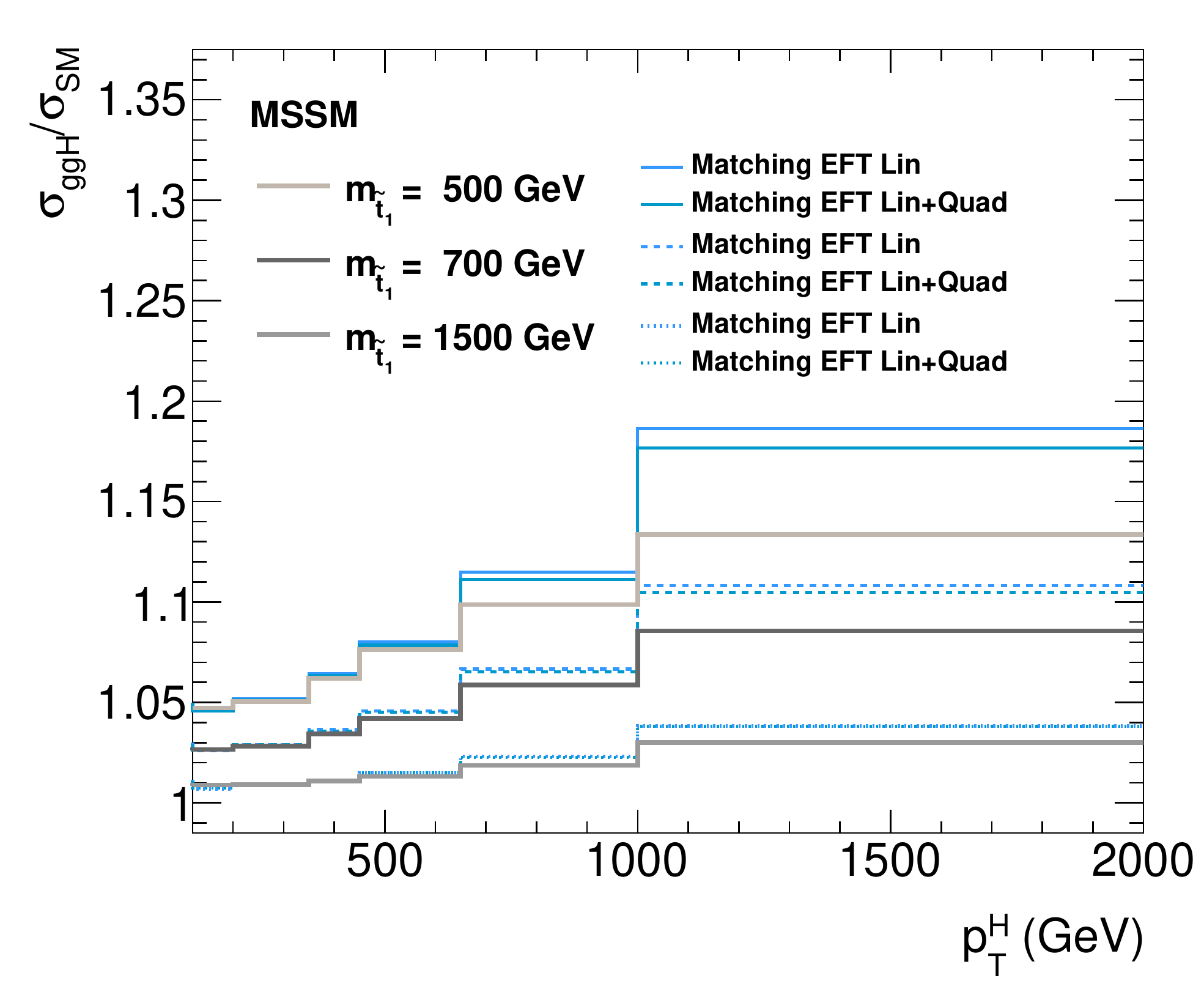}
    \caption{\label{fig:spectraMSSM}  \ptH spectra normalised to the SM prediction for the three MSSM benchmark points with varying $\tilde{t}_{1}$ masses. The SMEFT spectra for the matched Wilson coefficients and only linear and linear plus quadratic terms for the $m_{\tilde{t}}$=500\,GeV and 1500\,GeV benchmarks are also shown.}
  \end{center}
\end{figure}

\begin{figure}[t]
  \begin{center}
    \begin{tabular}{cc}
      \includegraphics[width=0.475\textwidth]{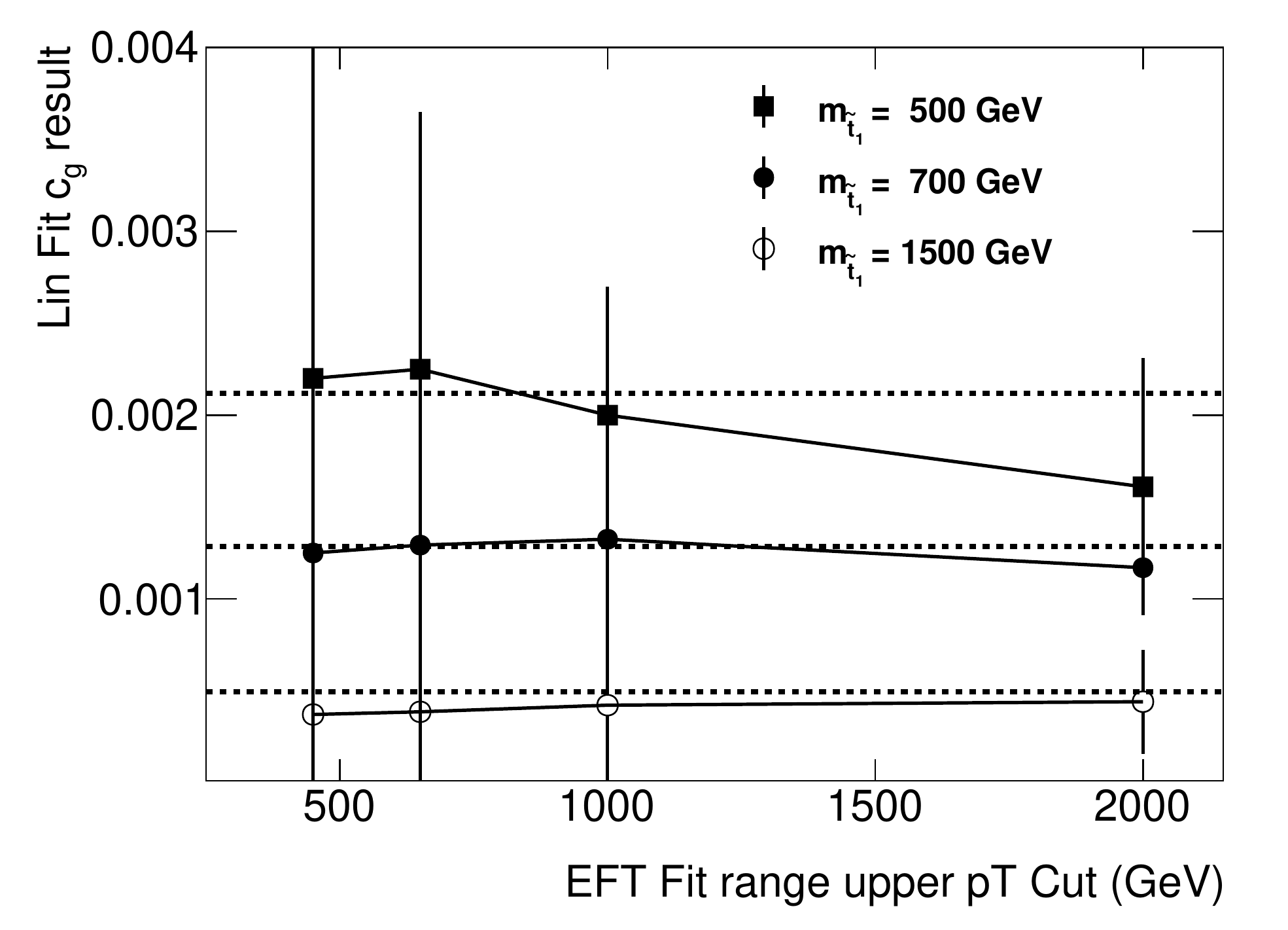} &
      \includegraphics[width=0.475\textwidth]{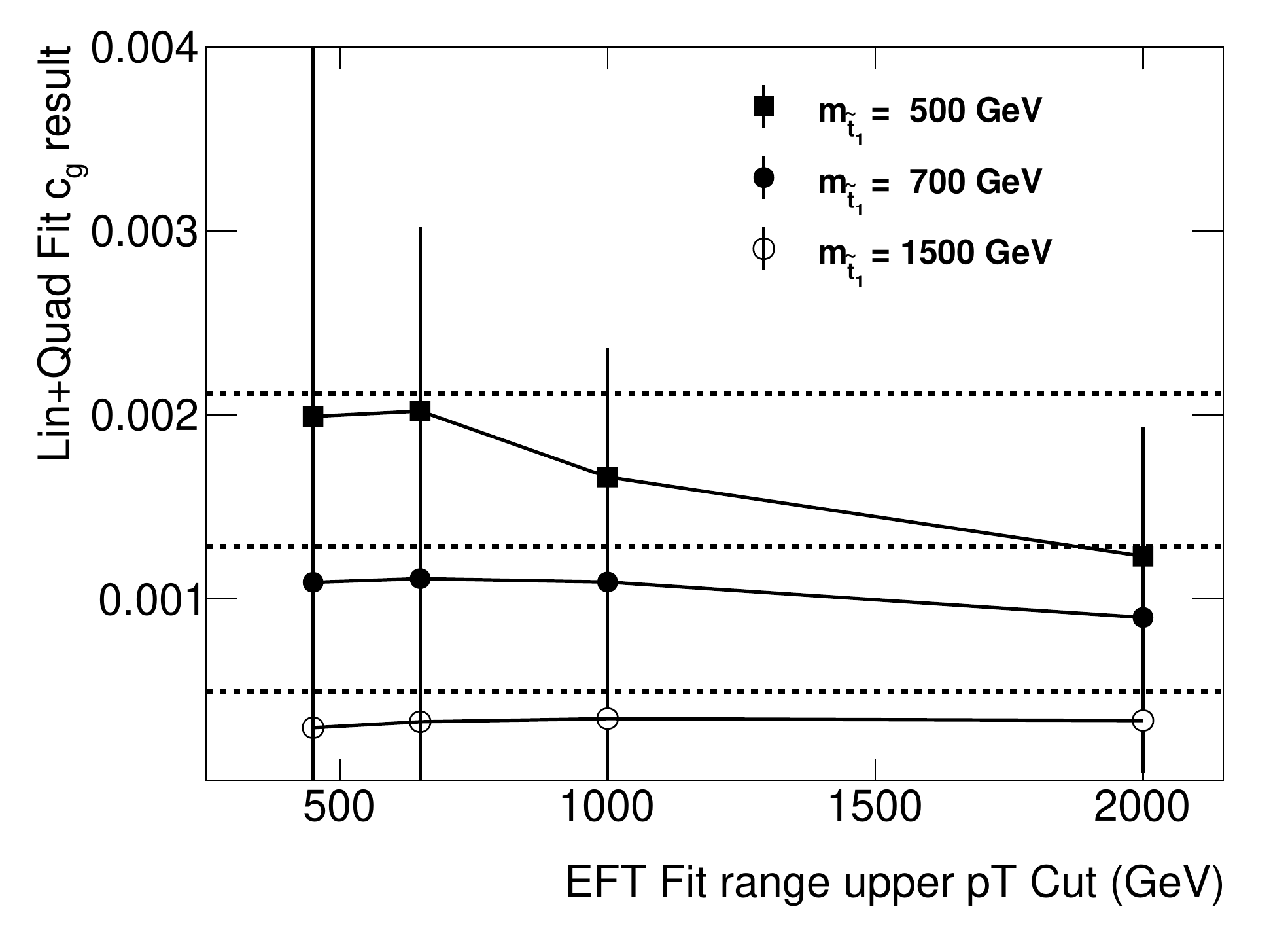} \\
    \end{tabular}
    \caption{\label{fig:fitMSSM} Wilson-coefficient $c_g$ values obtained with only linear (left) and with linear plus quadratic terms (right) to \ptH spectra produced from MSSM benchmark scenarios with different $m_{\tilde{t}_{1}}$ values as a function of the upper end of the \ptH fit range. The dashed lines show the $c_g$ values of the SMEFT matching to the MSSM benchmark scenarios.}
    \end{center}
\end{figure}

The MSSM \ptH{} spectra for different values of $m_{\tilde{t}_{1}}$ normalized to the SM predictions are compared in \fig{fig:spectraMSSM} to the corresponding SMEFT spectra for the $c_g$ values obtained from the matching relation of \eqn{eq:MSSMmatchcg}. Comparing \fig{fig:spectraHT} and \fig{fig:spectraMSSM} we see that the ratios $R_i$ for the MSSM benchmarks are much closer to unity than those observed for the heavy top partner model. This is due to the fact that the MSSM particles decouple in the high-mass limit.

Fits to these spectra are performed over a \ptH{} interval starting from 200\,GeV up to an upper limit ranging from 450\,GeV to 2\,TeV. The $c_g$ values extracted from the SMEFT fits are shown in \fig{fig:fitMSSM} as a function of the upper limit of the fits. Similarly to what has been observed for the heavy top partner model in the previous section, the fitted $c_g$ values are consistent with the matched values as long as the scalar top mass is significantly larger than the upper limit of the \ptH{} range used in the fit. When transverse-momentum values above the stop mass are used in the fit ($p_T^{H} \gtrsim m_{\tilde{t}_{1}}$), the fitted values of $c_g$ drop below the matched value of the model. It is interesting to observe how these trends indicate a sensitivity to New Physics from SMEFT fits beyond values of light stop masses already excluded by LHC direct searches \cite{CMS:2021eha,ATL-PHYS-PUB-2021-019}.

When comparing the results of SMEFT fits with linear and linear plus quadratic terms in \fig{fig:fitMSSM} we observe a similar dependence of the fitted $c_g$ values on the fit range in the linear SMEFT expansion as for the heavy top partner model. In the present case this dependence is milder because of the smaller values of $c_g$.
On the other hand, the central fit values obtained with the linear SMEFT are slightly closer to the matched values. This might appear in contradiction with the fact that in \fig{fig:spectraMSSM} the linear plus quadratic curves are always closer to the actual MSSM spectra. However, in the fits performed for the MSSM the spectrum normalisation is kept as a free parameter, which changes this picture. The relatively large uncertainties associated to the fitted $c_g$ values are a direct consequence of the decoupling nature of the MSSM scenario under consideration and the small relative corrections with respect to the SM. Therefore, both linear and linear plus quadratic SMEFT fits are compatible with the $c_g$ values obtained from the matching calculation within the range of applicability of the SMEFT, i.e.\ $p_T^{H} \lsim m_{\tilde{t}_{1}}$. This is a consequence of the decoupling behaviour of the stops and sbottoms as BSM particles at the dimension-6 level so that the linear expansion provides a reliable approximation for the differential cross section in this case.

\section{Results}
\label{sec:results}

This section presents the results of single- and multi-parameter SMEFT fits. The sensitivity of the LHC experiments to constrain the Wilson coefficients is studied by assuming a SM spectrum where the $R_i$ ratios in all bins are equal to unity. The uncertainties correspond to those published by the experiments for the LHC Run\,2 statistics and their extrapolations to the end of the HL-LHC programme at $3000$\,fb$^{-1}$ assuming a combination of ATLAS and CMS data. The fits are performed either by using directly the $R_i$ values from our scans and for the reference spectrum, thus being sensitive to both the spectrum shape and the Higgs signal yields normalised to the SM predictions, or by relying on the shape of the spectrum only, in which case only the relative change of the $R_i$ values as a function of \ptH are used in the fit. The sensitivity to constrain the Wilson coefficients of the two approaches is compared. The interplay of the $ggF$ and $t\bar tH$ production modes in the SMEFT fits is studied for the case of a SM-like spectrum and SMEFT benchmarks. The results presented in \sct{sec:ressens} include only the contribution of the $ggF$ process, while those in \sct{sec:procsens} also include $t\bar{t}H$ production.  Finally, in \sct{sec:resfitexp} the results reported by ATLAS and CMS are analysed in simultaneous fits to extract contours of the Wilson coefficients compatible with the current LHC data.

As mentioned above, constraints on the Wilson coefficients can also be obtained from the analysis of other processes and global fits have recently been performed. In particular, the study of $t \bar t$ and single top differential cross sections is sensitive to $c_{tg}$~\cite{Zhang:2010dr,Franzosi:2015osa}. The CMS collaboration has reported bounds on the chromomagnetic dipole operator from the $t \bar t$ differential cross section in the azimuthal angle between the two leptons in di-leptonic events~\cite{Sirunyan:2018ucr}. Other constraints have been obtained in \citere{CMS:2020lrr} by using top quark production with additional leptons. In the following we compare our results with the constraint on $c_{tg}$ coming from the global analysis in the top sector recently reported in \citere{Brivio:2019ius}, converted through Eq.~(\ref{eq:convctg}) to our conventions. 

\subsection{Experimental Results for Boosted Higgs Production}
\label{sec:resexp}

The ATLAS and CMS experiments have performed several analyses of Higgs production at high \ptH.
The CMS collaboration has conducted a search for Higgs bosons produced with \pt $>$ 450\,GeV decaying to a $b \bar b$ pair based on the Run\,2 data sample corresponding to an integrated luminosity of 137\,fb$^{-1}$~\cite{Sirunyan:2020hwz}. The search obtains an observed signal strength $\mu_H=3.7 \pm 1.6$ and reports an unfolded differential cross section in \ptH bins for $ggF$ production computed by assuming the other production modes at SM rates.

The ATLAS collaboration has reported preliminary results in a similar study for boosted Higgs production (inclusive over all production modes) in the $b \bar b$ final state on an integrated luminosity of 136\,fb$^{-1}$~\cite{ATLAS-CONF-2021-010}. The study reports a signal strengths of $\mu_H$ = 0.7 $\pm$ 3.3 and $\mu_H$ = 26 $\pm$ 31 in the fiducial regions defined by \ptH $\ge 450$\,GeV and \ptH $\ge 1$\,TeV, respectively, as well as unfolded inclusive signal strengths and differential cross sections in different \ptH bins.

ATLAS has also obtained results for the boosted $VH$ production process, again in the $H \rightarrow b \bar b$ decay channel, in the kinematic regions of 250 $< \ptV{} <$ 400\,GeV and \ptV{} $>$ 400\,GeV, where \ptV{} is the transverse momentum of the $W$ or $Z$ boson emitted with the Higgs boson~\cite{2021136204}. Finally, both collaborations have reconstructed the \ptH spectrum using the $H\to \gamma \gamma$ and $H\to Z Z\rightarrow 4 \ell$ decay channels~\cite{Sirunyan:2021rug,ATLAS-CONF-2019-032}.

These studies give a significant corpus of results that can be used to derive experimental relative uncertainties in the measurements of the Higgs yields in \ptH{} bins. The \ptH{} binning used by the experiments for inclusive boosted Higgs production is [300-450], [450-650], $\ge 650$\,GeV for CMS and [450-650], [650-1000], $\ge 1000$\,GeV for ATLAS. ATLAS and CMS have also provided estimates for the evolution of these uncertainties with 3\,ab$^{-1}$ of integrated luminosity at the HL-LHC~\cite{CMS-PAS-FTR-18-011,ATL-PHYS-PUB-2018-040}.

Our study is based on the latest results reported by ATLAS and CMS. At high \ptH{} we use the results of the boosted $H \rightarrow b \bar b$ analyses, while for \ptH{} values below those reported in the boosted analyses data 
are taken from the $H\to\gamma \gamma$ and $H\to Z Z\rightarrow 4 \ell$ analyses.

\subsection[{Sensitivity of the $ggF$ \ptH Spectrum to SMEFT Parameters}]{Sensitivity of the \boldmath{$ggF$} \boldmath{\ptH} Spectrum to SMEFT Parameters}
\label{sec:ressens}

In the following the sensitivity to constrain the Wilson coefficients from SMEFT fits for the $ggF$ process to the \ptH spectrum is studied by determining the regions in the SMEFT parameter space compatible with a SM-like spectrum, i.e.\ assuming $R_i$ ratios equal to unity in all \ptH{} bins.
When presenting results corresponding to an integrated luminosity of 140\,fb$^{-1}$, the bound on $c_{tg}$ reported in \citere{Brivio:2019ius} is shown as a blue-shaded band after translation to the conventions adopted in this paper. The translation of $c_{tg}$ includes the conversion and running according to \eqn{eq:convctg} and \eqn{eq:running}, respectively, assuming that \citere{Brivio:2019ius} uses an input value of $\mu_0=m_t$.

\begin{figure}[t]
  \begin{center}
    \begin{tabular}{cc}
      \includegraphics[width=0.45\textwidth]{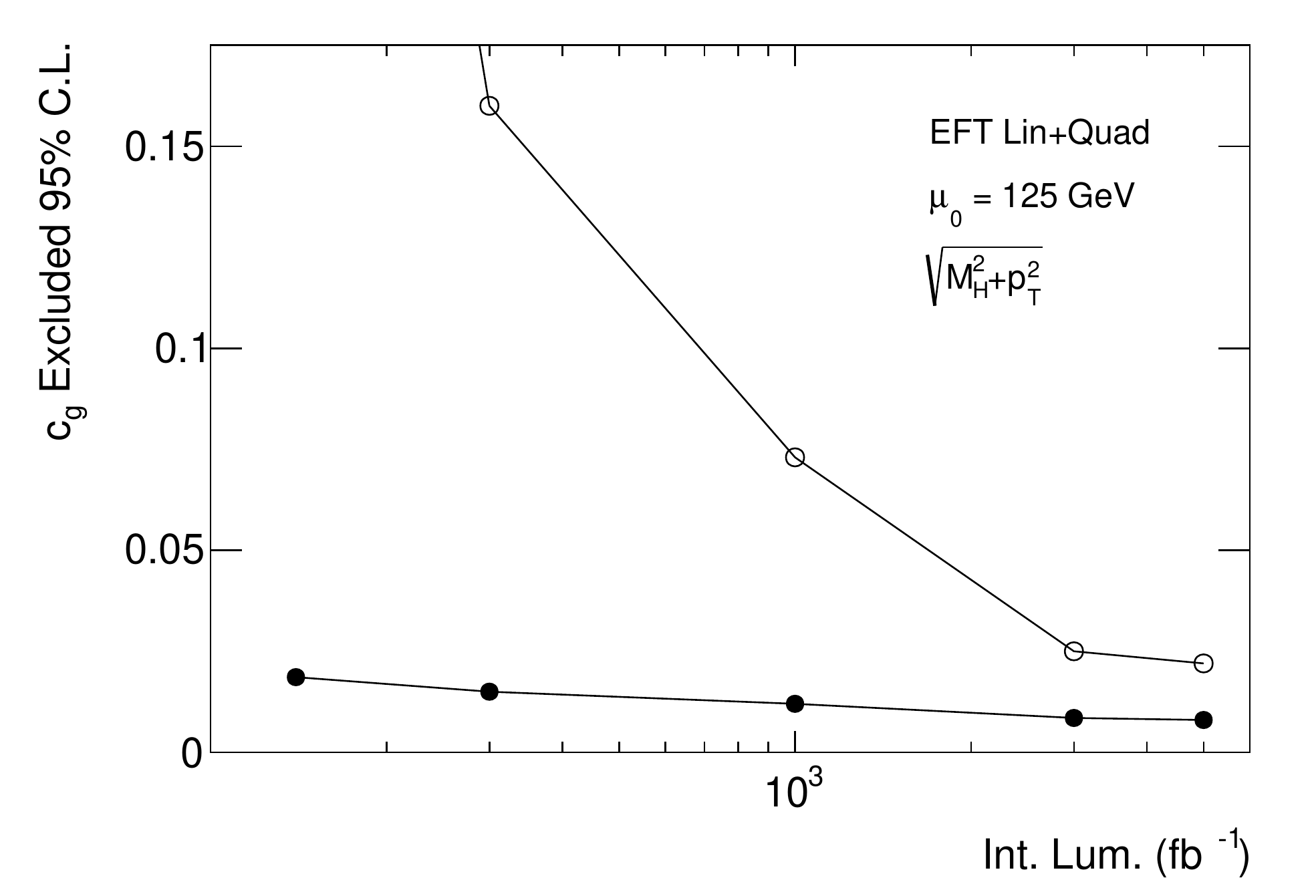} &
      \includegraphics[width=0.45\textwidth]{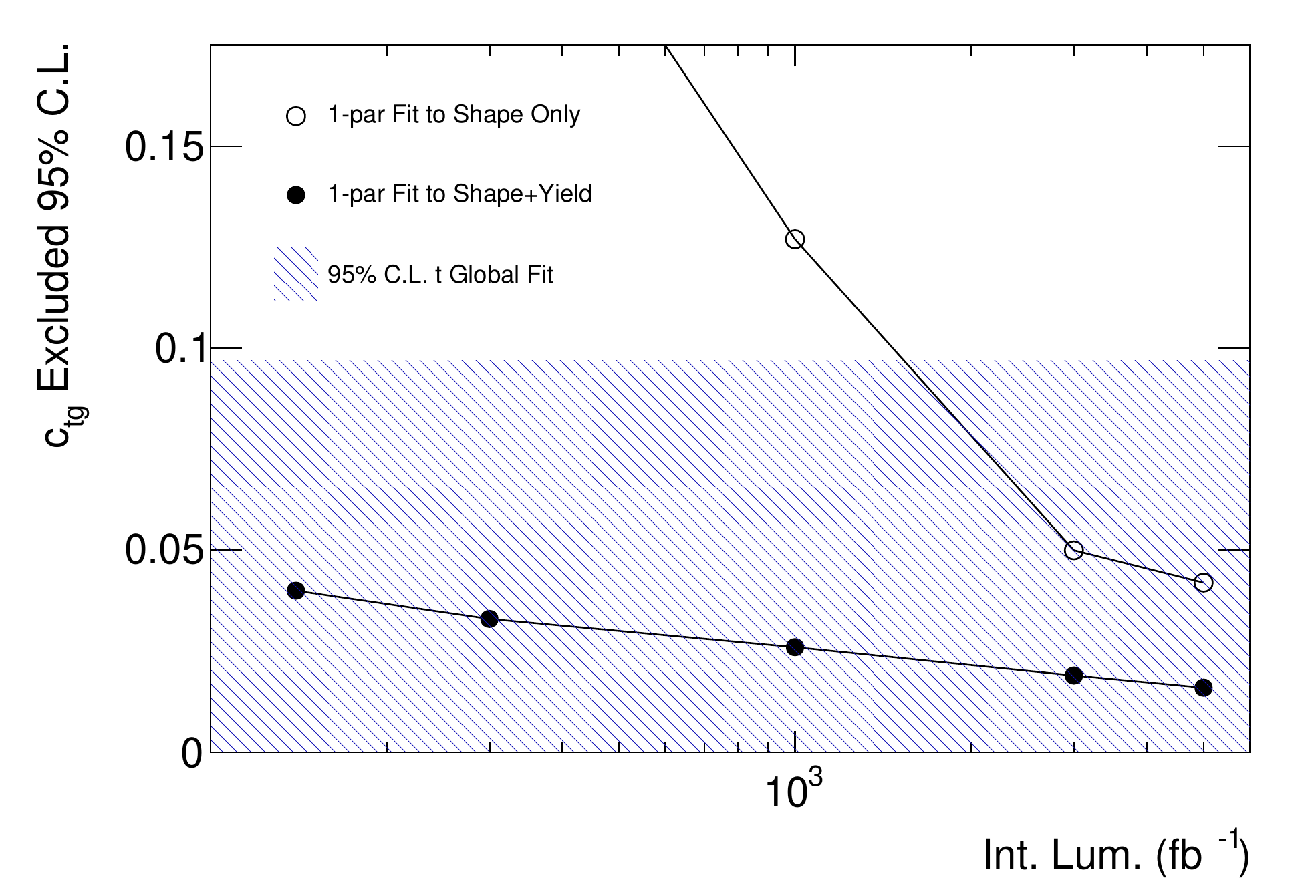} \\ 
    \end{tabular}
    \caption{\label{fig:exclcgctg} Values of the $c_g$ (left) and $c_{tg}$ (right) coefficients excluded at 95\%\,C.L. as a function of the integrated luminosity $\int {\cal{L}}$ for single-parameter fits. The open points refer to fits to the spectrum shape only and the filled point to fits including also the Higgs signal yield normalised to the SM. The shaded horizontal strip in the right plot indicates the 95\%\,C.L. interval for $c_{tg}$ from the current top quark fit of \citere{Brivio:2019ius} translated to our definition of $c_{tg}(\mu_0)$.}
    \end{center}
\end{figure}

First, we study the sensitivity to the $c_g$ and $c_{tg}$ coefficients as a function of the integrated luminosity $\int {\cal{L}}$ using the relative uncertainty on the determination of the Higgs $p_{T}$ spectrum for 140\,fb$^{-1}$ and 3000\,fb$^{-1}$ and scaling it as $1/\sqrt{\int{\cal{L}}}$ for other values.
This assumption is justified since the accuracy on the Higgs rate at high \ptH{} is dominated by the statistical uncertainty. \fig{fig:exclcgctg} shows the values of $c_g$ and $c_{tg}$ that can be excluded at 95\%\,C.L. in a single-parameter fit as a function of $\int {\cal{L}}$ by using either the spectrum shape only or also the signal yields in the fit, i.e.\ assuming $c_t(M_H)=1.0$. We see that the constraint from such one-parameter fit on $c_{tg}$ is already competitive with the constraint coming from top data \cite{Brivio:2019ius}.
These results demonstrate the significant improvement in accuracy achieved by adding the signal yield information, in particular when the available statistics is limited. The inclusion of LHC Run\,3 data is expected to provide us with a significant improvement of the strength of the constraints obtained from the study of boosted Higgs production.

\begin{figure}[t!]
  \begin{center}
    \begin{tabular}{cc}
      \includegraphics[width=0.425\textwidth]{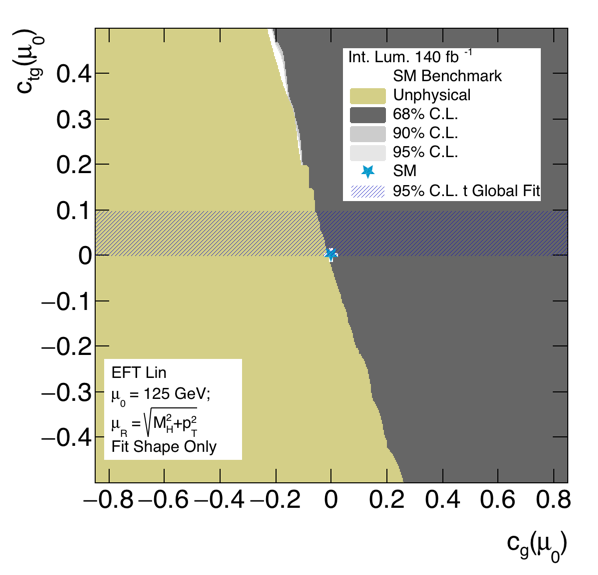} &
      \includegraphics[width=0.425\textwidth]{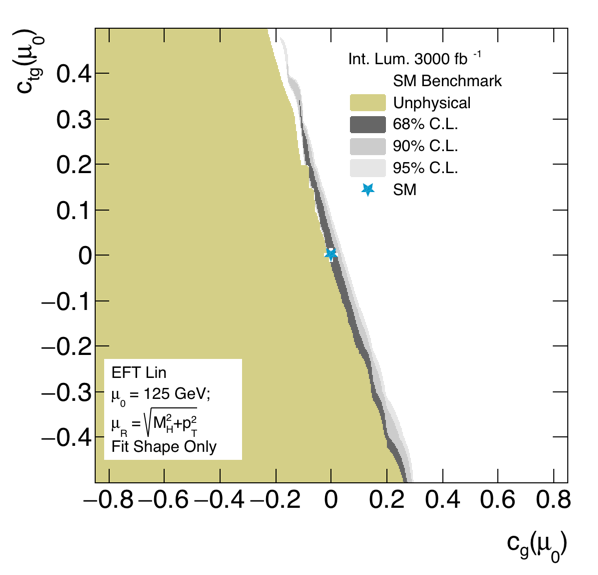} \\            
      \includegraphics[width=0.425\textwidth]{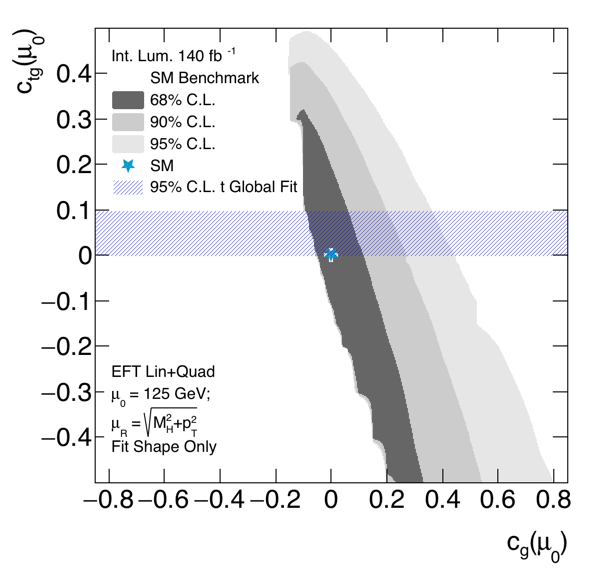} &

      \includegraphics[width=0.425\textwidth]{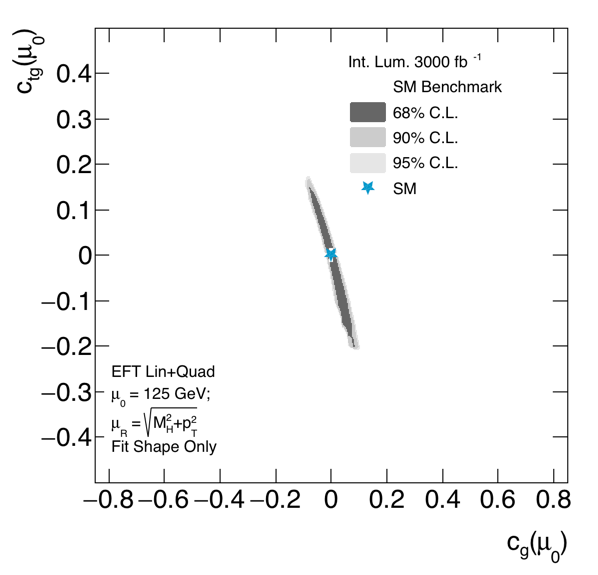} \\
    \end{tabular}
    \caption{\label{fig:cgctgnorm} Constraints on the $c_g$ and $c_{tg}$ coefficients at 68\% (dark grey), 90\% (mid grey) and 95\% (light grey) C.L. obtained from the fit to the spectrum shape on 200 $<\ptH<$ 2000\,GeV with 140\,fb$^{-1}$ (left panels) and 3000\,fb$^{-1}$ (right panels) for the linear (top panels) and with the addition of the quadratic terms (bottom panels). These constraints are obtained by including a free normalisation term in the fit and are therefore only sensitive to the spectrum shape. The SM value is indicated by the blue star and the best fit value by the white marker. The shaded horizontal strip indicates the 95\%\,C.L. interval for $c_{tg}$ from the top quark fit of \citere{Brivio:2019ius} translated to our definition of $c_{tg}(\mu_0)$. The region shown in colour and labelled ``unphysical'' in the fits with only SMEFT linear terms indicates the parameter space leading to negative unphysical values of cross section in the \ptH spectrum for a linear SMEFT approximation at dimension 6.}
    \end{center}
\end{figure}

\begin{figure}[t!]
  \begin{center}
    \begin{tabular}{cc}
      \includegraphics[width=0.425\textwidth]{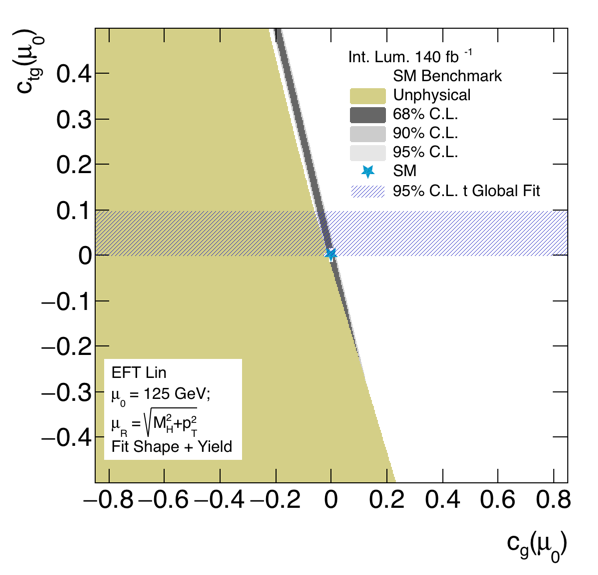} &
      \includegraphics[width=0.425\textwidth]{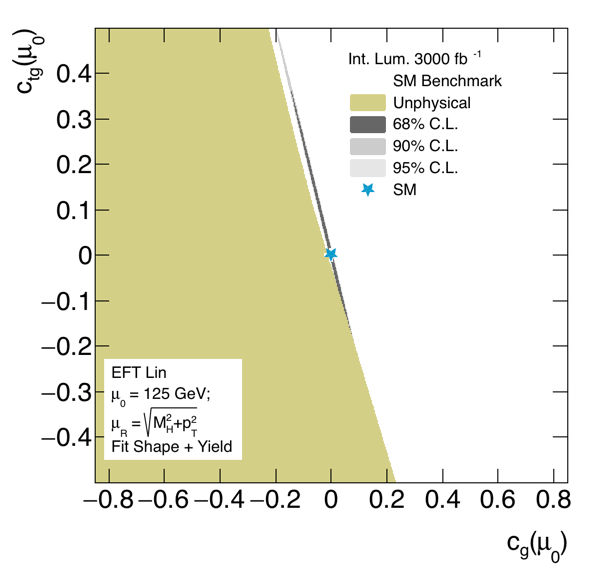} \\       
      \includegraphics[width=0.425\textwidth]{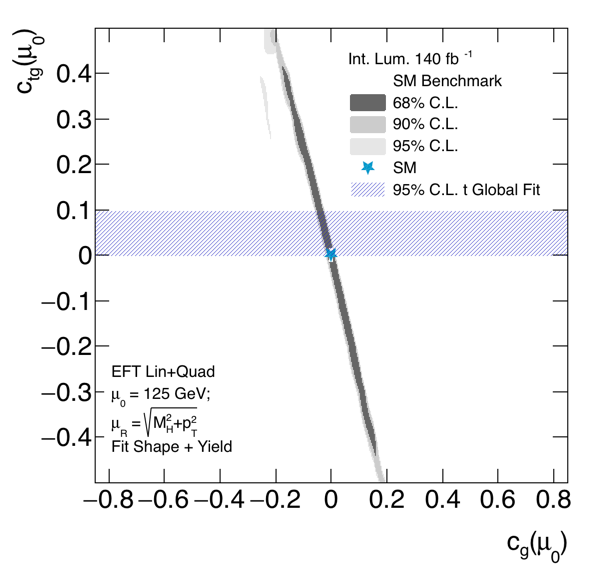} &
      \includegraphics[width=0.425\textwidth]{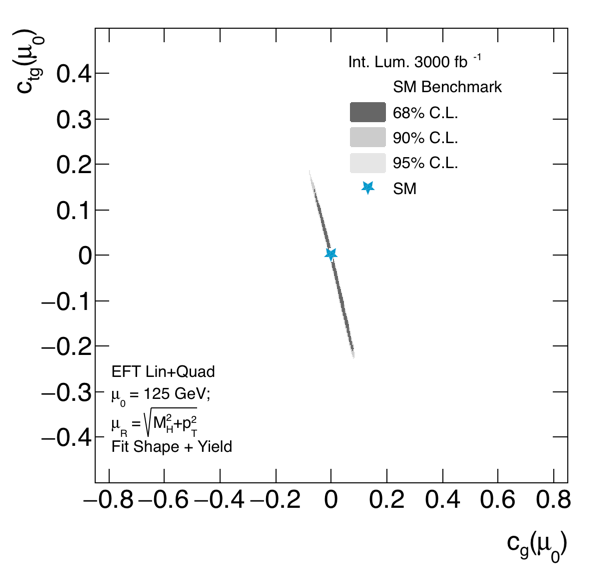} \\
    \end{tabular}
    \caption{\label{fig:cgctg} Constraints on the $c_g$ and $c_{tg}$ coefficients at 68\% (dark grey), 90\% (mid grey) and 95\% (light grey) C.L. obtained from the fit to the spectrum shape and Higgs signal rate on 200 $<\ptH<$ 2000\,GeV with 140\,fb$^{-1}$ (left panels) and 3000\,fb$^{-1}$ (right panels) for the linear (top panels) and with the addition of the quadratic terms (bottom panels). The SM value is indicated by the blue star and the best fit value by the white marker. The shaded horizontal strip indicates the 95\%\,C.L. interval for $c_{tg}$ from the top quark fit of \citere{Brivio:2019ius} translated to our definition of $c_{tg}(\mu_0)$. The region shown in colour and labelled ``unphysical'' in the fits with only SMEFT linear terms indicates the parameter space leading to negative unphysical values of cross section in the \ptH spectrum for a linear SMEFT approximation at \mbox{dimension 6.}}
    \end{center}
\end{figure}

\begin{figure}[t!]
  \begin{center}
    \begin{tabular}{cc}
      \includegraphics[width=0.425\textwidth]{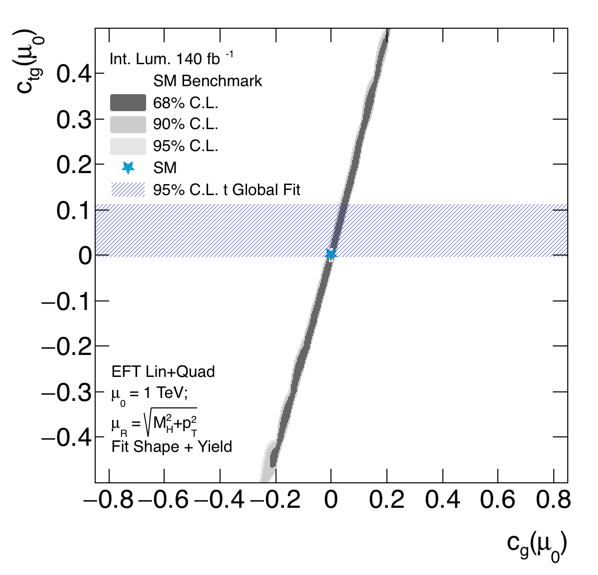} &
      \includegraphics[width=0.425\textwidth]{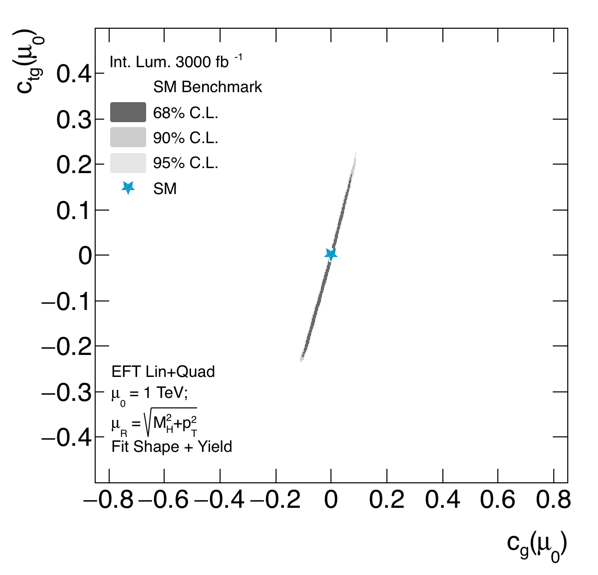} \\
    \end{tabular}
    \caption{\label{fig:cgctg1T} Constraints on the $c_g$ and $c_{tg}$ coefficients at 68\% (dark grey), 90\% (mid grey) and 95\% (light grey) C.L. obtained from the fit to the spectrum shape and Higgs signal rate on 200 $<\ptH<$ 2000\,GeV with 140\,fb$^{-1}$ (left panel) and 3000\,fb$^{-1}$ (right panel) for $\mu_0$ = 1\,TeV. The SM value is indicated by the blue star and the best fit value by the white marker. The shaded horizontal strip indicates the 95\%\,C.L. interval for $c_{tg}$ from the top quark fit of \citere{Brivio:2019ius} translated to our definition of $c_{tg}(\mu_0)$.}
    \end{center}
\end{figure}

We continue with simultaneous fits of $c_g$ and $c_{tg}$ in \figs{fig:cgctgnorm}--\ref{fig:cgctg-ggFttHBM1}.
The regions of the SMEFT parameter space compatible with a SM-like spectrum are determined for different scenarios and assumptions in the multi-parameter fits. Similar to what was done before, the sensitivity of the fits obtained by using only the spectrum shape and also including the signal yields are compared. We have verified that, for values consistent with the current constraints obtained by ATLAS~\cite{ATLAS-CONF-2020-027} and CMS~\cite{CMS-PAS-HIG-19-005} on the Higgs coupling modifier to top quarks, $\kappa_t$, in the $\kappa$-framework~\cite{Heinemeyer:2013tqa}, the $c_t$ coefficient does not induce significant modifications to the $\ptH$ spectrum shape and its contribution can be absorbed in a $c_t^2$ rescaling of the signal yield.

Furthermore, we compare the results obtained with a linear SMEFT expansion and an expansion including also quadratic terms in the Wilson coefficients.
As discussed in \sct{sec:theo} the default scale choice for our analysis is a dynamical factorization and renormalization scale $\mu_R = \mu_F = \sqrt{M_H^2+\pT^2}$.
As for the input scale $\mu_0$ for the Wilson coefficients, we adopt the natural choice $\mu_0$ = 125\,GeV, but, for comparison, we also show results for $\mu_0=1$\,TeV.

The experimental uncertainties are taken from the ATLAS and CMS Run\,2 public analyses (140\,fb$^{-1}$) and those expected for the HL-LHC with twenty times larger data sets (3000\,fb$^{-1}$). The relative uncertainties for the signal strengths in each \ptH{} bin are computed by combining the total uncertainties for ATLAS and CMS.  We also assume a theoretical uncertainty of 12\% in all \ptH{} bins.

\fig{fig:cgctgnorm} compares the sensitivity of the fits in the $c_{tg}$ -- $c_{g}$ plane 
for 140\,fb$^{-1}$ and 3000\,fb$^{-1}$ of integrated luminosity (left vs. right plot) as well as the difference between only linear against also quadratic terms in the SMEFT expansion (top vs. bottom plots). While in \fig{fig:cgctgnorm} only the shape information of the spectrum is used, the contours in \fig{fig:cgctg} are obtained from a fit that includes also the Higgs signal yields normalised to the SM. The results from the fits to the spectrum shapes and Higgs signal rates are not significantly modified if the assumption of $c_{t}(M_H)$ = 1.0 is relaxed by imposing instead a constraint on $c_{t}(M_H)$ corresponding to the $\kappa_t$ bounds reported by ATLAS~\cite{ATLAS-CONF-2020-027} and CMS~\cite{CMS-PAS-HIG-19-005}.

The use of SMEFT linear terms only can result in negative values of the Higgs cross section for some \ptH{} values in our fit range over a relatively large region of the $c_g$, $c_{tg}$ Wilson coefficients, which are marked as ``unphysical" in the respective figures. This is not unexpected and can be explained as an effect of the interference term between dimension-4 and dimension-6 operators, which can be negative, in absence of the squared contributions of the dimension-6 operators.

When comparing the top and bottom plots in  \fig{fig:cgctgnorm} and  \fig{fig:cgctg}, we observe that the inclusion of the SMEFT quadratic terms results in tighter constraints on the Wilson coefficients. Whether quadratic terms should or should not be included in dimension-6 SMEFT fits is still an open question: including quadratic terms corresponds to promote the SMEFT expansion from the level of the Lagrangian to the amplitude level. As we have seen in \sct{sec:rangeHT} and \sct{sec:rangeMSSM}, the issue is also related to the decoupling behaviour of the underlying UV model.

The effects from including quadratic terms are particularly significant in \fig{fig:cgctgnorm} where the fits include a free normalisation parameter and are therefore only sensitive to the spectrum shape and not to the Higgs signal yields normalised to the SM. Comparing the left and right panels in \fig{fig:cgctgnorm} and  \fig{fig:cgctg}, it is also clear that with the statistical accuracy at $3000$\,fb$^{-1}$ the constraints will substantially improve and also become significantly more stable with respect to the various choices that can be made in the fits. Disregarding the case of the shape-only fit with linear terms (upper left plot in \fig{fig:cgctgnorm}) for which the sensitivity obtained at 140~fb$^{-1}$ is still limited, we observe a significant anti-correlation between the $c_g$ and $c_{tg}$ coefficients defining elongated contours where they are compatible with a SM-like spectrum. 

Up to now all our results have been obtained by using a dynamical renormalisation and factorisation scale $\mu_R = \mu_F=\sqrt{M_H^2+\pT^2}$. In the case in which all the scales are fixed to $\mu_F=\mu_R=\mu_0=M_H$ the constraints obtained for a SM-like spectrum would be tighter by $\simeq$10\%. This, however, is originating from an ill-defined treatment of logarithmic contributions at higher orders and should not be viewed as an improvement.

Finally, we consider the case in which the reference scale $\mu_0$ is chosen as $\mu_0=1$\,TeV, instead of $\mu_0=125$\,GeV, still using a dynamical scale $\mu_R = \mu_F=\sqrt{M_H^2+\pT^2}$. The corresponding results are shown in \fig{fig:cgctg1T}. Comparing with \fig{fig:cgctg}, we observe that the correlation between $c_{tg}$ and $c_g$ is reversed. This is due to the fact that the RG equation for $c_g$ has a large term driven by $c_{tg}$, see \eqn{eq:running}. However, the two results are fully compatible and can be directly obtained from each other by translating each point in the parameter space for one $\mu_0$ scale choice to the other through the RG evolution of the Wilson coefficients.
The last two results underline the importance of including the RG running of the Wilson coefficients and properly specifying the input scale of the fit.

\subsection[Interplay of $ggF$ and $t\bar tH$ Production in the \ptH SMEFT Fits]{Interplay of \boldmath{$ggF$} and \boldmath{$t\bar tH$} Production in the \boldmath{\ptH{}} SMEFT Fits}
\label{sec:procsens}

\begin{figure}[b!]
  \begin{center}
    \begin{tabular}{cc}
      \includegraphics[width=0.425\textwidth]{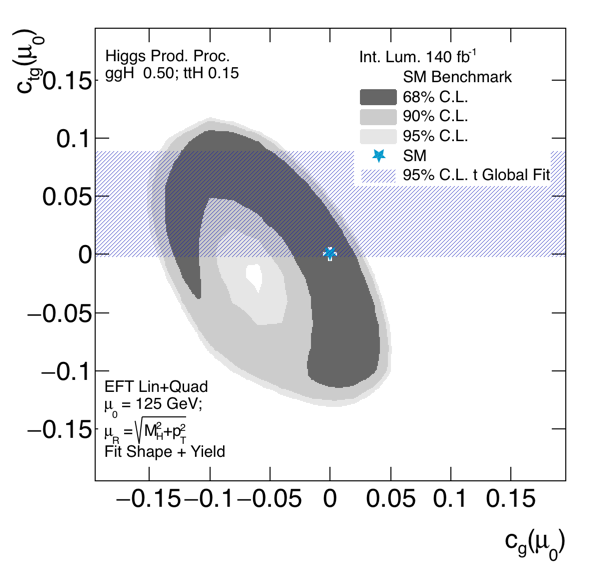} &
      \includegraphics[width=0.425\textwidth]{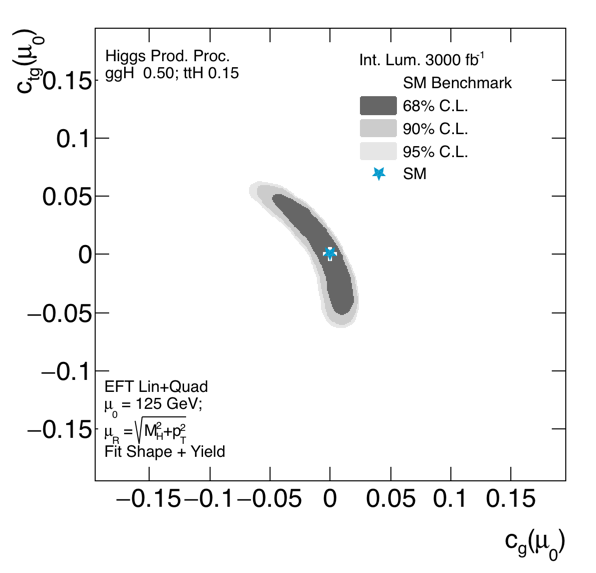} \\
    \end{tabular}
    \caption{\label{fig:cgctg-ggFttHSM} Constraints on the $c_g$ and $c_{tg}$ coefficients at 68\% (dark grey), 90\% (mid grey) and 95\% (light grey) C.L. obtained from the fit including $ggF$ and $t\bar tH$ SMEFT contributions to the SM-like spectrum shape and Higgs signal rate on 200 $<\ptH<$ 2000\,GeV with 140\,fb$^{-1}$ (left panel) and 3000\,fb$^{-1}$ (right panel). The SM value is indicated by the blue star and the best fit value by the white marker. The shaded horizontal strip indicates the 95\%\,C.L. interval for $c_{tg}$ from the top quark fit of \citere{Brivio:2019ius} translated to our definition of $c_{tg}(\mu_0)$. Notice also the reduced range compared to \fig{fig:cgctg}.}
    \end{center}                  
\end{figure}      

The results presented so far are based on the study of the SMEFT effects in Higgs production through gluon fusion. Although these effects are dominant, as discussed in \sct{sec:prod}, in general the experimental analyses select a mixture of boosted Higgs events from all production modes. In particular, the $ggF$ and $t\bar tH$ production processes are sensitive to the same SMEFT operators. Therefore, it is interesting to study the effect of including both $ggF$ and $t\bar tH$ contributions in the SMEFT fits of the Wilson coefficients. Again we consider a SM-like spectrum where the $R_i$ ratios are equal to unity in all \ptH{} bins with the relative uncertainties taken from the experimental measurements and projections. Additionally, benchmark scenarios with non-zero $c_g$ and/or $c_{tg}$ values are fitted, assuming the same relative uncertainties. As discussed in \sct{sec:fit}, the simultaneous fit of the $ggF$ and $t\bar tH$ SMEFT spectra is based on the sum of the $R_i$ ratios for each process weighted by the respective fraction, $f$, of signal events in the sample. For the $VH$ and VBF processes, whose spectra are not sensitive to the ${\cal O}_1$, ${\cal O}_2$ and ${\cal O}_3$ SMEFT operators, a SM-like contribution is assumed.       

First, the regions in the $c_g$ -- $c_{tg}$ parameter space compatible with a SM-like spectrum are determined assuming $f_{ggF}$ = 0.50 and $f_{t{\bar t}H}$ = 0.15, consistent with the fractions of signal events obtained by the experimental analyses. The results are shown in \fig{fig:cgctg-ggFttHSM} for 140\,fb$^{-1}$ and 3000\,fb$^{-1}$, including linear and quadratic terms as well as \ptH{} shape and signal yield information in the fit for $\mu_0=125$\,GeV.         
Comparing the results in \fig{fig:cgctg-ggFttHSM} to those in the corresponding two lower plots of \fig{fig:cgctg}, the additional correlation between the two Wilson coefficients, induced by the not negligible sensitivity offered by the $t\bar tH$ process to $c_{tg}$, is evident, in particular when $c_g$ is small. For example, in the case of 140\,fb$^{-1}$, we see that the extreme values of the allowed region in the $ggF$-only fit ($c_{tg}\sim \pm\,0.4$, $c_g\sim\pm\, 0.2$) are now completely excluded. Indeed, the strong (anti-)correlation of $c_g$ and $c_{tg}$ resulting from the $ggF$-only fit is largely reduced yielding a more circular shape in the $c_g$ -- $c_{tg}$ plane.

\begin{figure}[t]
  \begin{center}
    \begin{tabular}{cc}
      \includegraphics[width=0.425\textwidth]{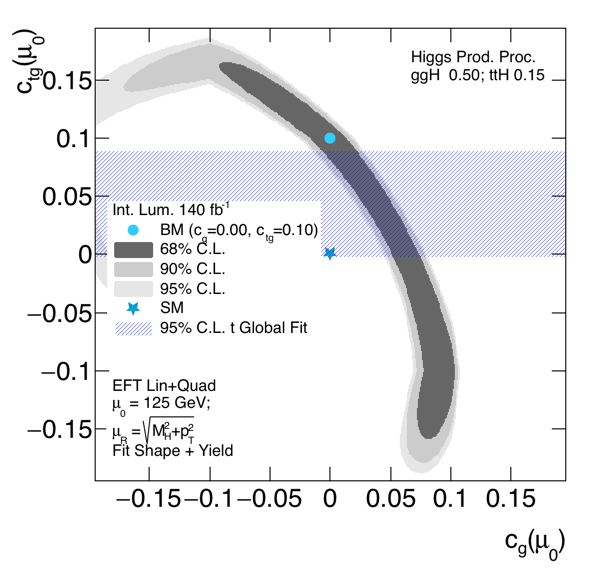} &
      \includegraphics[width=0.425\textwidth]{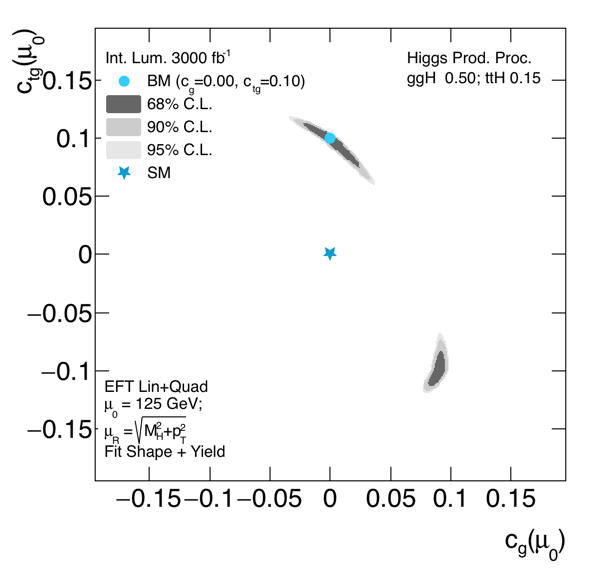} \\
    \end{tabular}
    \caption{\label{fig:cgctg-ggFttHBM1} Constraints on the $c_g$ and $c_{tg}$ coefficients at 68\% (dark grey), 90\% (mid grey) and 95\% (light grey) C.L. obtained from the fit including $ggF$ (0.50) and $t\bar tH$ (0.15) SMEFT contributions to the spectrum shape and Higgs signal rate on 200 $<\ptH<$ 2000\,GeV with 140\,fb$^{-1}$ (left panel) and 3000\,fb$^{-1}$ (right panel) for a benchmark point with $c_g(\mu_0)=0$ and $c_{tg}(\mu_0)=0.1$. The benchmark value is indicated by the dark blue circle. The shaded horizontal strip indicates the 95\%\,C.L. interval for $c_{tg}$ from the top quark fit of \citere{Brivio:2019ius} translated to our definition of $c_{tg}(\mu_0)$.}
    \end{center}                  
\end{figure}

We continue by considering benchmark spectra generated by combining SM-normalised SMEFT distributions for the $ggF$ and $t{\bar t}H$ processes corresponding to a set of Wilson coefficients. The values of $c_g$, $c_{tg}$ compatible with these benchmark spectra are determined by fitting them either with templates for the consistently combined $ggF$ and $t\bar tH$ predictions or by using only $ggF$ predictions, while assuming all the other Higgs production processes to be SM-like. \fig{fig:cgctg-ggFttHBM1} shows the constraints obtained for a benchmark with $c_g(\mu_0)=0$ and $c_{tg}(\mu_0)=0.10$ fitted with the same $ggF$ (0.50) and $t{\bar t}H$ (0.15) SMEFT contributions used to generate the benchmark spectrum. This point in the parameter space is chosen at $c_{g}$=0 and at the upper bound on $c_{tg}$ obtained by the fit to top-quark data of \citere{Brivio:2019ius} where the $t{\bar t}H$ effects are largest. The $ggF$ plus $t\bar tH$ fit correctly recovers the input benchmark values. Instead, if this spectrum is fitted assuming the $t\bar tH$ contribution to be SM-like, the fit yields $c_g(\mu_0) = -0.04\pm 0.07$ and $c_{tg}(\mu_0) = 0.17\pm 0.33$ for 140\,fb$^{-1}$ and $c_g(\mu_0) = -0.04\pm 0.01$ and $c_{tg} = 0.19\pm 0.04$ for 3000\,fb$^{-1}$ of integrated luminosity. Similarly, a benchmark $c_g(\mu_0)=0.02$ and $c_{tg}(\mu_0)=0.08$ gives fit values of $c_g(\mu_0) = -0.02\pm 0.01$ and $c_{tg}(\mu_0) = 0.14\pm 0.03$ for 3000\,fb$^{-1}$ of integrated luminosity.
This indicates that the assumption of a SM-like $t\bar tH$ contribution in the SMEFT fit gives results that are still statistically compatible with the benchmark true parameters at the current level of accuracy, but these would become significantly biased with the uncertainties anticipated for the HL-LHC.  We conclude that a global fit including the $ggF$ and $t\bar tH$ SMEFT contributions weighted by the corresponding fractions of signal events selected by the experimental analysis will be required for HL-LHC analyses.

\subsection[Fit of the $ggF$ Spectrum to Experimental Results]{Fit of the \boldmath{$ggF$} Spectrum to Experimental Results}
\label{sec:resfitexp}

\begin{figure}[b!]
  \begin{center}
    \includegraphics[width=0.675\textwidth]{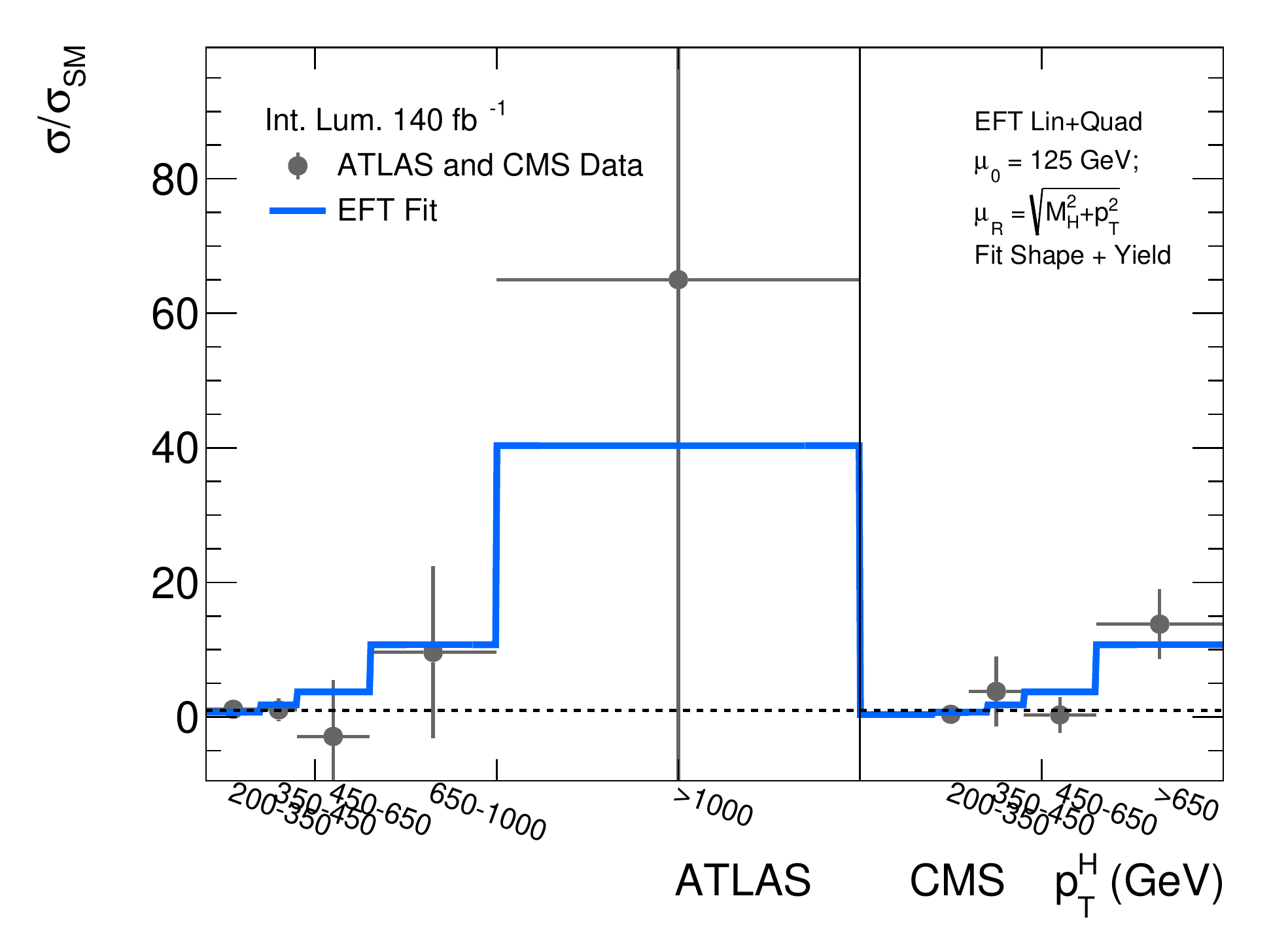} 
    \caption{\label{fig:fitcombspectrum} Result of a simultaneous fit to the spectrum shape and Higgs signal rate of ATLAS (left) and CMS (right) using the SMEFT expansion with linear and quadratic terms at $\mu_0=125$\,GeV. The points with error bars show the data and the continuous line the best fit corresponding to $c_{g} (M_H) =-0.08$ and $c_{tg} (M_H) =-0.13$.}
  \end{center}
\end{figure}

The ATLAS and CMS collaborations have already published constraints on the SMEFT parameters obtained by fits to their \ptH results~\cite{Sirunyan_2019,ATLAS-CONF-2020-053}. However, the early ATLAS analysis for $ggF$ production at high \ptH{} of \citere{ATLAS-CONF-2020-053} only used the $H \rightarrow \gamma \gamma$ results for \ptH $\ge$ 450\,GeV in their fits, while the CMS SMEFT result of \citere{Sirunyan_2019} was based on the preliminary boosted $H \rightarrow b \bar b$ analysis performed with only 35\,fb$^{-1}$ of integrated luminosity. Here we provide a first assessment of the values of the Wilson coefficients compatible with the current LHC results from both experiments, while consistently accounting for their RG evolution.
Since no combination is available of the \ptH{} spectra measured by ATLAS and CMS, we obtain constraints from the data of the two experiments by performing a simultaneous SMEFT fit to the two independent sets of signal strenghts in the \ptH{} bins reported by the two experiments. The uncertainties on these measurements are largely dominated by the statistical contributions, correlations between the two results are expected to be small at the current level of accuracy and are neglected here.

\begin{figure}[t]
  \begin{center}
    \begin{tabular}{cc}
    \includegraphics[width=0.475\textwidth]{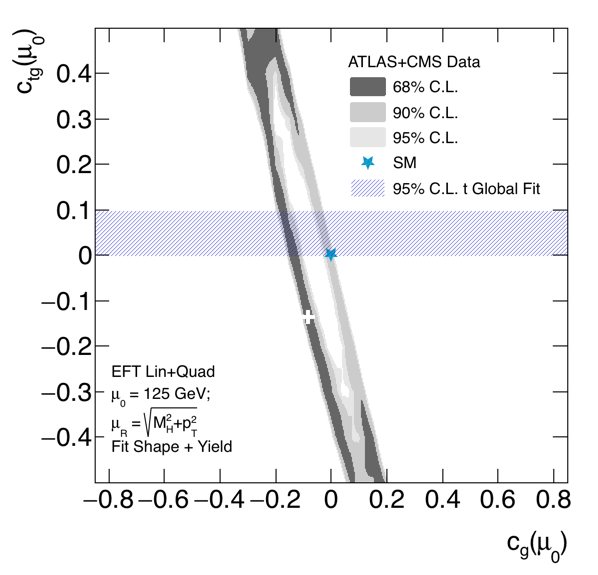} &
    \includegraphics[width=0.475\textwidth]{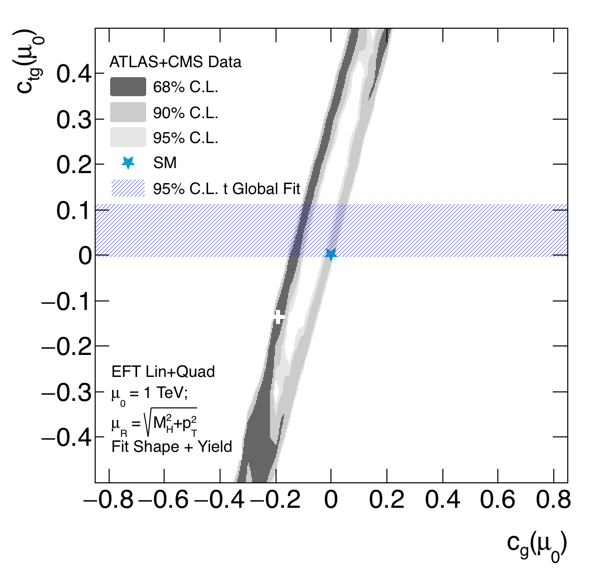} \\ 
    \end{tabular}
    \caption{\label{fig:fitcombcgctg} Constraints on the $c_g$ and $c_{tg}$ coefficients at 68\% (dark grey), 90\% (mid grey) and 95\% (light grey) C.L. obtained from the simultaneous fit to ATLAS and CMS data for $\mu_0 = 125$\,GeV (left panel) and $\mu_0 = 1$\,TeV (right panel). The SM value is indicated by the blue star and the best fit value by the white marker. The shaded horizontal strip indicates the 95\%\,C.L. interval for $c_{tg}$ from the top quark fit of \citere{Brivio:2019ius} translated to our definition of $c_{tg}(\mu_0)$.}
  \end{center}
\end{figure}

The ATLAS preliminary analysis is inclusive, i.e.\ the signal strengths are not separated by Higgs production modes. The CMS analysis reports results for $ggF$ signal strengths assuming the other production modes to be SM-like. Therefore, we consider only the $ggF$ SMEFT contribution in our study, and the ATLAS signal strength values and related uncertainties are rescaled by the inverse of the relative $ggF$ contribution reported by ATLAS in each \ptH{} bin.

The fit is performed by keeping the $c_g$ and $c_{tg}$ parameters free. Given the limited accuracy of the current experimental results only the fits using both the spectrum shape and the signal event rate as well as including both linear and quadratic SMEFT terms offer some sensitivity to the Wilson coefficients, see \sct{sec:ressens}. Therefore our fit does not include the normalisation parameter, i.e.\ assuming $c_t(M_H)=1.0$, and it uses linear and quadratic terms in the SMEFT expansion. The results of a simultaneous fit to the preliminary ATLAS and the CMS data is shown in \fig{fig:fitcombspectrum} for the best fit values $c_g (M_H) =-0.08$ and $c_{tg} (M_H) =-0.13$. The corresponding contraints in the $c_{tg}$ -- $c_g$ plane obtained for two choices of the RG input scale, $\mu_0=125$\,GeV and $\mu_0=1$\,TeV, are summarised in \fig{fig:fitcombcgctg} with data compatible with the SM at better than 90\% C.L.

The fit yields a contour in the parameter space compatible with the data within 90\%\,C.L that can be described as a very elongated ellipse with an elliptic hole in the middle. This essentially leads to two disjoint sets of solutions along almost parallel lines. The SM point lies close to the edge of one of these lines, while the best fit value belongs to the other parallel branch. These features are fully preserved when changing the input scale $\mu_0$ from 125\,GeV to 1\,TeV. However, as observed in \fig{fig:cgctg1T}, the use of the reference scale $\mu_0=1$\,TeV leads to a striking modification of the correlation in the $c_{tg}$ -- $c_{g}$ plane. As discussed in \sct{sec:ressens}, this effect is due to the large impact of the $c_{tg}$ term in the RG equation of the $c_g$ coefficient, see \eqn{eq:running}. The two results can be consistently translated into each other by using the RG evolution of the operators.

\section{Conclusions}
\label{sec:concl}

In this paper we have considered Higgs boson production at large transverse momentum, a new stage in the study of the Higgs profile at the LHC.  The effects of the three leading dimension-6 operators in the gluon-fusion and $t{\bar t}H$ production processes have been studied in the Standard Model Effective Field Theory. They include a point-like Higgs--gluon
coupling, the chromomagnetic dipole operator of the top quark and a reweighting 
of the top-quark Yukawa coupling. The corresponding Wilson coefficients $c_g$, $c_{tg}$ and $c_t$ have been defined as scale dependent quantities obeying renormalization group equations coupled among the three operators.

The central goal of our study has been to place bounds on the Wilson coefficients of the dimension-6 operators from current LHC data and to assess the sensitivity that can be reached throughout the LHC programme. This has been achieved through multi-parameter $\chi^2$ fits of the deviations of the Higgs production cross section computed in SMEFT in bins of Higgs $p_T$ combined with the available state-of-the-art SM predictions for the $ggF$ and $t{\bar t}H$ channels, taking into account the experimental and theoretical uncertainties.

Although not yet critical at the current level of experimental accuracy, accounting for renormalisation group evolution of the SMEFT operators is crucial when it comes to precision bounds on the Wilson coefficients. Our analysis employs a dynamical renormalisation and factorisation scale, which is set to the transverse mass of the Higgs boson for the $ggF$ production mode. To account for the correct momentum transfer in the SMEFT couplings the dynamical setting of the renormalization scale is used also for the Wilson coefficients. 
In the fits the Wilson coefficients are then extracted at an input scale $\mu_0$ equal to the Higgs mass. We have shown that the use of a high scale $\mu_0=1$ TeV would lead to a completely different correlation between $c_g$ and $c_{tg}$ 
when performing multi-parameter fits in the $c_{tg}$ -- $c_g$ plane. With an analysis spanning a wide range of Higgs transverse momenta, running effects are important. It is therefore advisable to present results including the running of the operators as this will become essential for HL-LHC analysis.
 
The validity of the SMEFT assumptions has been tested by performing fits to a model with a heavy-top partner and to the MSSM with a light scalar top, varying the new particle masses and the \ptH range used in the fits. The Wilson coefficients extracted from the fits agree with the values obtained by calculating the matching between the explicit model and the SMEFT as long as the upper end of the \ptH range used in the fits is not significantly larger than the mass of the heavy particle in the model. We have also contrasted the results obtained with only linear or linear plus quadratic terms in the SMEFT expansion. In the case in which new physics does not decouple, as it happens in scenarios with a heavy top partner,  SMEFT effects are large and a fit with linear and quadratic terms yields results that are closer to those obtained in the full model. Instead, when the new physics decouples, as in the case of the MSSM with a light scalar top and heavy SUSY particles, the new physics effects are small and can be captured already through a fit including only linear terms in the  SMEFT expansion.

The sensitivity to the $c_g$ and $c_{tg}$ Wilson coefficients has been studied for both the current accuracy and for future LHC accuracies of boosted Higgs production. The use of both the shape of the \ptH{} spectrum and the signal yield provide stringent constraints on the Wilson coefficients, with a peculiar (anti-)correlation in the $c_{tg}$ -- $c_g$ plane. The constraint on $c_{tg}$ extracted from a single-parameter fit to a SM-like Higgs \ptH{} spectrum appears to be competitive with the corresponding bound obtained from a multi-parameter fit in the global analysis of the top sector recently reported in \citere{Brivio:2019ius}. The constraints in the $c_{tg}$ -- $c_g$ plane will become rather tight with the anticipated accuracy obtained with HL-LHC data.

In the range allowed by the current ATLAS and CMS constraints on the Higgs coupling modifier to top quarks, $\kappa_t$, the $c_t$ Wilson coefficient does not induce significant modifications to the $\ptH$ spectrum shape. Moreover, it is clear that constraints on $c_t$ will come from inclusive cross section measurements, especially from $t\bar t H$ production, while the Higgs transverse-momentum spectrum will then yield an accurate determination of the Wilson coefficients $c_g$ and $c_{tg}$.

Most of our results have been obtained by focusing on the $ggF$ production mode. However, we have also investigated the interplay between the $ggF$ and $t{\bar t}H$ Higgs production, which are both sensitive to the same set of Wilson coefficients, assuming the other Higgs production modes to be SM-like. The SMEFT effects from $t{\bar t}H$ production are subdominant and its inclusion does not lead to substantial effects for current uncertainties, with the exception of the region at low $c_g$ and large $c_{tg}$ values where the effects from $t{\bar t}H$ production are largest. These effects are not yet observable in the current LHC analyses, but will need to be included as the experimental accuracy will improve. Moreover, we stress that with future uncertainties the combination of $ggF$ and $t{\bar t}H$ in SMEFT fits will 
enable the reduce the correlation of $c_g$ and $c_{tg}$ and lead to stringent constraints on both of them simultaneously.

 Finally, we have extracted constraints on the $c_{g}$ and $c_{tg}$ for $ggF$ production from a simultaneous fit to the preliminary ATLAS and the CMS data and compared our results with those obtained in a global SMEFT analysis in the top sector.  With the current data the ATLAS and CMS results are compatible with the SM at better than 90\% C.L.

\vskip 0.5cm
\noindent {\bf Acknowledgements}
 
\noindent We are grateful to Andrea Sciandra for his contribution in the early stages of this work, to Eleni Vryonidou for useful correspondence and to Hannah Arnold for her comments. M.B.\ thanks the Galileo Galilei Institute for Theoretical Physics, Arcetri, Florence, for hospitality and support during part of the preparation of this study. The work of M.G.\ is supported in part by the Swiss National Foundation under contract 200020$\_$188464.
 
\appendix

\section{Running of the Wilson Coefficients}
\label{sec:appendix}
In this appendix we describe the solution of the renormalization-group
equations (RGEs) at the leading-log level. We have extracted the RGEs
from \citere{Deutschmann:2017qum} and confirmed the leading-log part by
the corresponding results of \citeres{Jenkins:2013zja,
Jenkins:2013wua, Alonso:2013hga}. In order to translate the
notation of \citere{Deutschmann:2017qum} into our notation of
\eqn{eq:OPs}, we used the following relations,
\begin{eqnarray}
C_1 & = & -(c_t-1) \frac{\sqrt{2}m_t}{v}~\frac{\Lambda^2}{v^2}\,, \nonumber \\
C_2 & = & \frac{c_g}{8\pi^2}~\frac{\Lambda^2}{v^2} \,,\nonumber \\
C_3 & = & c_{tg} \frac{m_t}{\sqrt{2}{v}}~\frac{\Lambda^2}{v^2}\,,
\label{eq:translation}
\end{eqnarray}
where $v = 1/\sqrt{\sqrt{2}G_F}$ denotes the vacuum expectation value of
the Higgs field and $m_t$ the top mass, while $\Lambda$ represents the
cut-off scale of the SMEFT framework. The parameters $m_t$ and the
Wilson coefficients $c_t, c_g$ and $c_{tg}$ evolve either due to strong
interactions or due to the top-Yukawa coupling in our QCD analysis. The relevant RGEs for
our work at the leading-log level are ($a_s = \alpha_s/\pi$),
\begin{eqnarray}
\partial_t C_1 & = & -a_s~C_1 + 8
\frac{m_t^2}{v^2}~a_s~C_3\,, \nonumber \\
\partial_t C_2 & = & \frac{m_t}{8\sqrt{2} \pi^2 v}~C_3\,, \nonumber \\
\partial_t C_3 & = & \frac{a_s}{6}~C_3\,,
\label{eq:rge0}
\end{eqnarray}
where $\partial_t = \partial/\partial t$, with $t=\log(Q^2/\mu^2)$ with $Q$
being the scale of the physical process and $\mu$ the input scale of the
RG-evolution. The corresponding RGEs for our parameters can be derived
from \eqn{eq:rge0},
\begin{eqnarray}
\partial_t c_t & = & -4 \frac{m_t^2}{v^2}~a_s~c_{tg} \,,\nonumber \\
\partial_t c_g & = & \frac{m_t^2}{2 v^2}~c_{tg} \,,\nonumber \\
\partial_t c_{tg} & = & \frac{7a_s}{6}~c_{tg}\,.
\label{eq:rge}
\end{eqnarray}
The simplicity of these RGEs underlines the suitability of our conventions in \eqn{eq:OPs}, i.e.\ the leading QCD evolution is factored out, leaving us with pure EFT effects on the scale dependence in \eqn{eq:rge}. If $c_{tg}$ vanishes at the input scale $\mu$, there is no scale dependence of our Wilson coefficients $c_t, c_g, c_{tg}$ at LL-level. For the SM parameters, we use the 5-flavour scheme,
i.e.\ all effects at the scale of the top mass and beyond are decoupled from the running in compatibility with the definition of the strong
coupling $\alpha_s$ in the PDF fits,
\begin{eqnarray}
\partial_t m_t & = & -a_s~m_t\,, \nonumber \\
\partial_t a_s & = & -\beta_0 a_s^2\,,
\end{eqnarray}
with $\beta_0 = (33-2N_F)/12 = 23/12$ denoting the leading-order beta
function coefficient of QCD. The second RGE for $a_s$ can be used
as usual to replace the integration measure $t$ by an integration over
$a_s$,
\begin{equation}
dt = -\frac{d a_s}{\beta_0 a_s^2}\,.
\end{equation}
This allows us to solve the RGEs for the running top mass and the Wilson
coefficient $C_3$ immediately at LL level,
\begin{eqnarray}
m_t(Q^2) & = & m_t(\mu^2) \left(\frac{a_s(Q^2)}{a_s(\mu^2)}
\right)^{\frac{1}{\beta_0}} \,,\nonumber \\
C_3(Q^2) & = & C_3(\mu^2) \left(\frac{a_s(Q^2)}{a_s(\mu^2)}
\right)^{-\frac{1}{6\beta_0}}\,.
\end{eqnarray}
The solution for $C_3$ corresponds to the following LL
expression for $c_{tg}$,
\begin{equation}
c_{tg}(Q^2) = c_{tg}(\mu^2) \left(\frac{a_s(Q^2)}{a_s(\mu^2)}
\right)^{-\frac{7}{6\beta_0}}\,.
\end{equation}
Using the solutions for $m_t$ and $c_{tg}$, the RGEs for $c_t$ and $c_g$
can be solved,
\begin{eqnarray}
c_t(Q^2) & = & c_t(\mu^2) +
\frac{24}{5}~\frac{m_t^2(\mu^2)}{v^2}~c_{tg}(\mu^2)~\left\{
\left(\frac{a_s(Q^2)}{a_s(\mu^2)} \right)^{\frac{5}{6\beta_0}} - 1
\right\}\,, \nonumber \\
c_g(Q^2) & = & c_g(\mu^2) -
\frac{3}{5-6\beta_0}~\frac{m_t^2(\mu^2)}{v^2}~
\frac{c_{tg}(\mu^2)}{a_s(\mu^2)}~\left\{
\left(\frac{a_s(Q^2)}{a_s(\mu^2)} \right)^{\frac{5}{6\beta_0}-1} - 1
\right\}\,.
\end{eqnarray}
In order to cope with pure QCD effects beyond the LL level,
we are adding the next-to-leading QCD part to the RGEs of $a_s$ and
$c_g$,
\begin{eqnarray}
\partial_t a_s & = & -\beta_0 a_s^2 - \beta_1 a_s^3 \,,\nonumber \\
\partial_t c_g & = & \frac{m_t^2}{2 v^2}~c_{tg} - \beta_1 a_s^2 c_g\,,
\end{eqnarray}
where $\beta_1 = (153-19 N_F)/24 = 29/12$ denotes the NLO coefficient of
the QCD beta function. The final solution of the RG running for $c_g$ can then be extended approximately as
\begin{equation}\label{eq:cgrun}
c_g(Q^2) = \frac{\beta_0 + \beta_1 a_s(Q^2)}{\beta_0 + \beta_1
a_s(\mu^2)} \left\{ c_g(\mu^2) -
\frac{3}{5-6\beta_0}~\frac{m_t^2(\mu^2)}{v^2}~
\frac{c_{tg}(\mu^2)}{a_s(\mu^2)}~\left[
\left(\frac{a_s(Q^2)}{a_s(\mu^2)} \right)^{
\frac{5}{6\beta_0}-1} - 1 \right] \right\}\,,
\end{equation}
where the error is of next-to-leading-logarithmic order for the top
Yukawa-induced contributions, while the leading QCD part agrees with the
known scale dependence of the trace-anomaly coefficient
\cite{Callan:1970yg, Symanzik:1970rt, Coleman:1970je, Crewther:1972kn,
  Chanowitz:1972vd, Chanowitz:1972da}. As seen from \eqn{eq:cgrun}, when $c_{tg}$ vanishes only the pure QCD running of $c_g$ remains.

\bibliography{higgsfit}

\end{document}